\documentclass[final]{elsart}

\usepackage{natbib}

\usepackage{graphicx}

\usepackage{amssymb}

\journal{New Astronomy}

\usepackage{txfonts}
\usepackage{relsize}
\usepackage{rotate}

\usepackage{exscale}

\begin{document}

\newcommand{\labeln}[1]{\label{#1}}
\newcommand{\Msolar}{M$_{\odot}$}
\newcommand{\Lsolar}{L$_{\odot}$}
\newcommand{\farcmin}{\hbox{$.\mkern-4mu^\prime$}}
\newcommand{\farcsec}{\hbox{$.\!\!^{\prime\prime}$}}
\newcommand{\kms}{\rm km\,s^{-1}}
\newcommand{\cc}{\rm cm^{-3}}
\newcommand{\Alfven}{$\rm Alfv\acute{e}n$}
\newcommand{\Vap}{V^\mathrm{P}_\mathrm{A}}
\newcommand{\Vat}{V^\mathrm{T}_\mathrm{A}}
\newcommand{\D}{\partial}
\newcommand{\DD}{\frac}
\newcommand{\TAW}{\tiny{\rm TAW}}
\newcommand{\mm }{\mathrm}
\newcommand{\Bp }{B_\mathrm{p}}
\newcommand{\Bpr }{B_\mathrm{r}}
\newcommand{\Bpz }{B_\mathrm{\theta}}
\newcommand{\Bt }{B_\mathrm{T}}
\newcommand{\Vp }{V_\mathrm{p}}
\newcommand{\Vpr }{V_\mathrm{r}}
\newcommand{\Vpz }{V_\mathrm{\theta}}
\newcommand{\Vt }{V_\mathrm{\varphi}}
\newcommand{\Ti }{T_\mathrm{i}}
\newcommand{\Te }{T_\mathrm{e}}
\newcommand{\rtr }{r_\mathrm{tr}}
\newcommand{\rbl }{r_\mathrm{BL}}
\newcommand{\rtrun }{r_\mathrm{trun}}
\newcommand{\thet }{\theta}
\newcommand{\thetd }{\theta_\mathrm{d}}
\newcommand{\thd }{\theta_d}
\newcommand{\thw }{\theta_W}
\newcommand{\beq}{\begin{equation}}
\newcommand{\eeq}{\end{equation}}
\newcommand{\ben}{\begin{enumerate}}
\newcommand{\een}{\end{enumerate}}
\newcommand{\bit}{\begin{itemize}}
\newcommand{\eit}{\end{itemize}}
\newcommand{\barr}{\begin{array}}
\newcommand{\earr}{\end{array}}
\newcommand{\DroII}{\overline{\overline{\rm D}}}
\newcommand{\DroI}{{\overline{\rm D}}}
\newcommand{\eps}{\epsilon}
\newcommand{\veps}{\varepsilon}
\newcommand{\vepsdi}{{\cal E}^\mathrm{d}_\mathrm{i}}
\newcommand{\vepsde}{{\cal E}^\mathrm{d}_\mathrm{e}}
\newcommand{\lraS}{\longmapsto}
\newcommand{\lra}{\longrightarrow}
\newcommand{\LRA}{\Longrightarrow}
\newcommand{\Equival}{\Longleftrightarrow}
\newcommand{\DRA}{\Downarrow}
\newcommand{\LLRA}{\Longleftrightarrow}
\newcommand{\diver}{\mbox{\,div}}
\newcommand{\grad}{\mbox{\,grad}}
\newcommand{\cd}{\!\cdot\!}
\newcommand{\Msun}{{\,{\cal M}_{\odot}}}
\newcommand{\Mstar}{{\,{\cal M}_{\star}}}
\newcommand{\Mdot}{{\,\dot{\cal M}}}
\newcommand{\ds}{ds}
\newcommand{\dt}{dt}
\newcommand{\dx}{dx}
\newcommand{\dr}{dr}
\newcommand{\dth}{d\theta}
\newcommand{\dphi}{d\phi}

\newcommand{\pt}{\frac{\partial}{\partial t}}
\newcommand{\pk}{\frac{\partial}{\partial x^k}}
\newcommand{\pj}{\frac{\partial}{\partial x^j}}
\newcommand{\pmu}{\frac{\partial}{\partial x^\mu}}
\newcommand{\pr}{\frac{\partial}{\partial r}}
\newcommand{\pth}{\frac{\partial}{\partial \theta}}
\newcommand{\pR}{\frac{\partial}{\partial R}}
\newcommand{\pZ}{\frac{\partial}{\partial Z}}
\newcommand{\pphi}{\frac{\partial}{\partial \phi}}

\newcommand{\vadve}{v^k-\frac{1}{\alpha}\beta^k}
\newcommand{\vadv}{v_{Adv}^k}
\newcommand{\dv}{\sqrt{-g}}
\newcommand{\fdv}{\frac{1}{\dv}}
\newcommand{\dvr}{{\tilde{\rho}}^2\sin\theta}
\newcommand{\dvt}{{\tilde{\rho}}\sin\theta}
\newcommand{\dvrss}{r^2\sin\theta}
\newcommand{\dvtss}{r\sin\theta}
\newcommand{\dd}{\sqrt{\gamma}}
\newcommand{\ddw}{\tilde{\rho}^2\sin\theta}
\newcommand{\mbh}{M_{BH}}
\newcommand{\dualf}{\!\!\!\!\left.\right.^\ast\!\! F}
\newcommand{\cdt}{\frac{1}{\dv}\pt}
\newcommand{\cdr}{\frac{1}{\dv}\pr}
\newcommand{\cdth}{\frac{1}{\dv}\pth}
\newcommand{\cdk}{\frac{1}{\dv}\pk}
\newcommand{\cdj}{\frac{1}{\dv}\pj}
\newcommand{\rad}{\;r\! a\! d\;}

\begin{frontmatter}


%

\title{An implicit numerical algorithm for solving the general relativistic
hydrodynamical equations around accreting compact objects}
\author{A. Hujeirat\corauthref{cor}},
\corauth[cor]{Corresponding author.
Tel.: ++49 6221 5417 63, fax: +49 6221 5417 02, 
Address: ZAH, Landessternwarte, K\"onigstuhl 12, D-69117 Heidelberg, Deutschland}
\ead{A.Hujeirat@lsw.uni-heidelberg.de}
\author{M. Camenzind} \and
\ead{M.Camenzind@lsw.uni-heidelberg.de}
\author{B.W. Keil}
\ead{B.Keil@lsw.uni-heidelberg.de}
\address{ZAH, Landessternwarte Heidelberg-K\"onigstuhl,
  Universit\"at Heidelberg, 69117 Heidelberg, Germany}

\begin{abstract}

An implicit algorithm  for solving the equations of general
relativistic hydrodynamics in conservative form in three-dimensional
axi-symmetry is presented. This algorithm is a direct extension of
the pseudo-Newtonian implicit radiative magnetohydrodynamical solver
-IRMHD- into the general relativistic regime.

We adopt  the Boyer-Lindquist coordinates and formulate the
hydrodynamical equations in the fixed background of a Kerr black
hole. The set of equations are solved implicitly using the
hierarchical solution scenario (HSS).  The HSS is efficient, robust
and enables the use of a variety of solution procedures that range
from a purely explicit up to fully implicit schemes. The
discretization of the HD-equations is based on the finite volume
formulation and the defect-correction iteration strategy for
recovering higher order
spatial and temporal accuracies. 
Depending on the astrophysical problem, a variety of relaxation
methods can be applied. In particular the vectorized black-white
Line-Gauss-Seidel relaxation method is most suitable for modeling
accretion flows with shocks, whereas the Approximate Factorization
Method is for weakly compressible flows.

The results of several test calculations that verify  the accuracy
and robustness of the algorithm are shown. 
Extending the algorithm to enable solving
the non-ideal MHD equations in the general relativistic regime is the subject of
our ongoing research.

\end{abstract}

\begin{keyword}
plasmas \sep (magnetohydrodynamics:) MHD \sep gravitation \sep
relativity \sep shock waves \sep methods: numerical
\PACS 95.30.Qd \sep 95.30.Sf
\end{keyword}

\end{frontmatter}



\section{Introduction}

The field of astrophysical fluid dynamics (henceforth AFD) deals
with the macroscopic evolution of gaseous-matter and plasmas in a
wide variety of circumstances in astrophysics. The scope of AFD is
broad, encompassing topics such as star formation, accretion
phenomena onto compact and young stellar objects, dynamics of the
interstellar medium, jets, winds and outflows emerging from around
young stellar objects, from quasars and microquasars, supernovae
explosion,
$\gamma-$ray bursts and structure formation in the universe.

One of the ultimate aims of numerical astrophysics is to develop a
black box algorithm which contains numerical solvers that are
unconditionally stable, robust, efficient, Newtonian, relativistic
and capable of treating flows that are strongly compressible, weakly
incompressible, self-gravitating, radiating, magnetized
multi\--com\-po\-nent-plasmas with high spatial and temporal accuracies on
unstructured meshes and to provide the required solutions within the
scale of hours. While this goal is unlikely to be achieved within
the next few years, the increasing number of sophisticated numerical
algorithms developed
during the last two decades is remarkably encouraging. 
In particular, significant improvements have been achieved in
increasing the spatial dimensions and
enhancing  the efficiency and accuracy of numerical algorithms \citep[see e.g.][]{Nagel_etal2006}. \\
On the other hand, the problem of robustness of the solvers in AFD
has been barely considered  nor even seriously discussed. 
In this paper we discuss the robustness problem in AFD, present
enhancement strategies and address the necessity of constructing
robust general relativistic implicit and radiative MHD solvers.
For completeness we review the concepts of efficiency and robustness of numerical
solvers in computational fluid dynamics.

A numerical solver is said to be relatively efficient if the
corresponding number of algebraic manipulations per time step per
computer-processor can be made respectively small. As a consequence,
using  high performance computers with large number of computing
processors, a significant progress has been achieved in improving
runtime efficiencies,  provided the computing load is distributed
appropriately. Thus, by means of modern hardware technology, the
efficiency of computer-codes could be enhanced even without
modifying the employed mathematical approach. This  has led to
uncoordinated developments of relatively large number of
computer-codes, that may differ in efficiency and integrated
physical processes but are almost essentially identical with respect
to the mathematical solution procedure. This phenomenon can be
attributed to the lack of robustness. By robustness, we mean the
capability of a solver to be applied to a large class of problems
without
modifying the algorithmic core significantly.

\begin{figure}[htp]
\begin{center}
\hspace*{-0.08 \hsize}
\includegraphics*[angle=90, width=0.40 \hsize, bb=0 0 313 450,clip ]{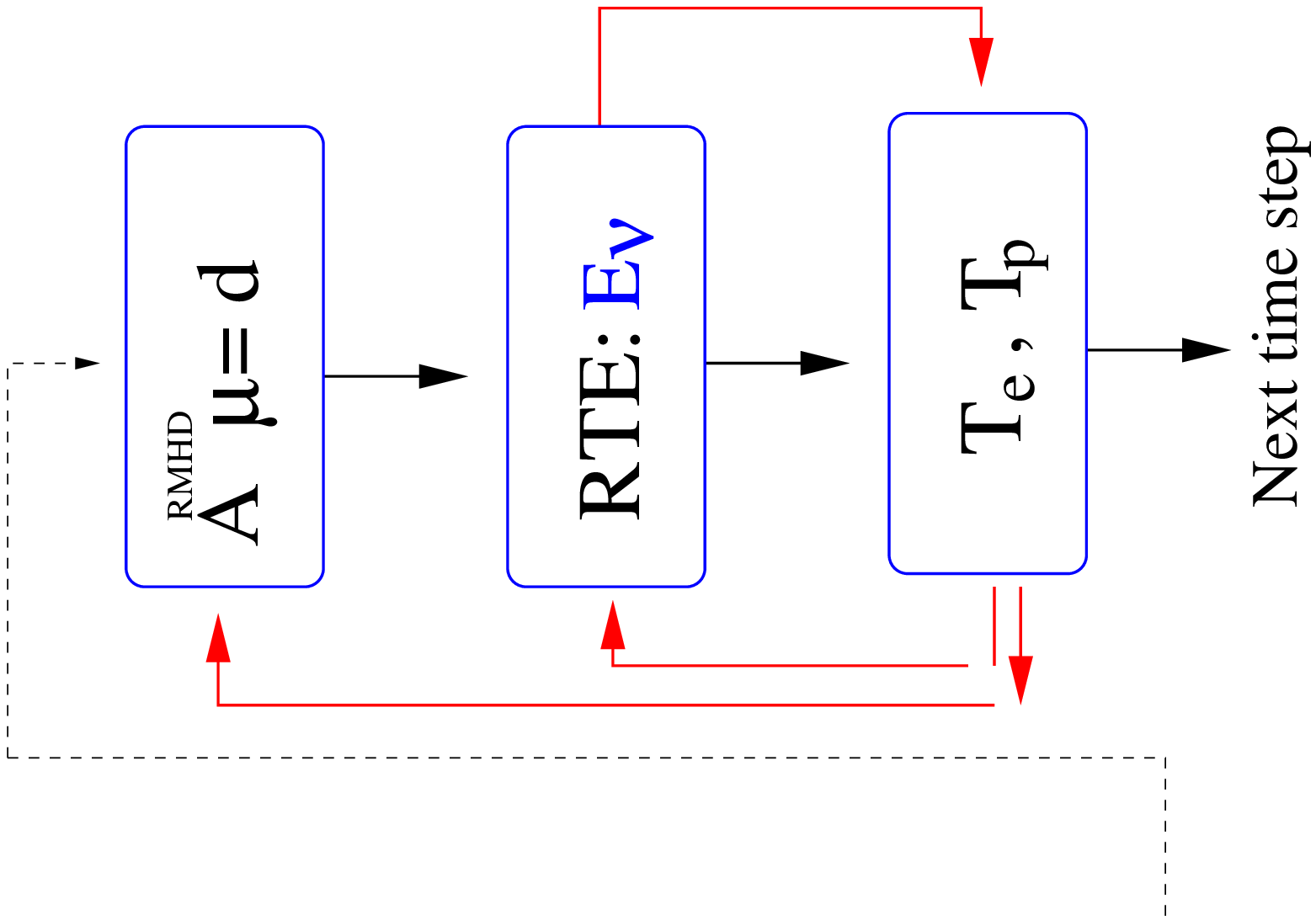} \\
\includegraphics*[angle=90, width=0.86 \hsize, bb=0 0 498 460,clip ]{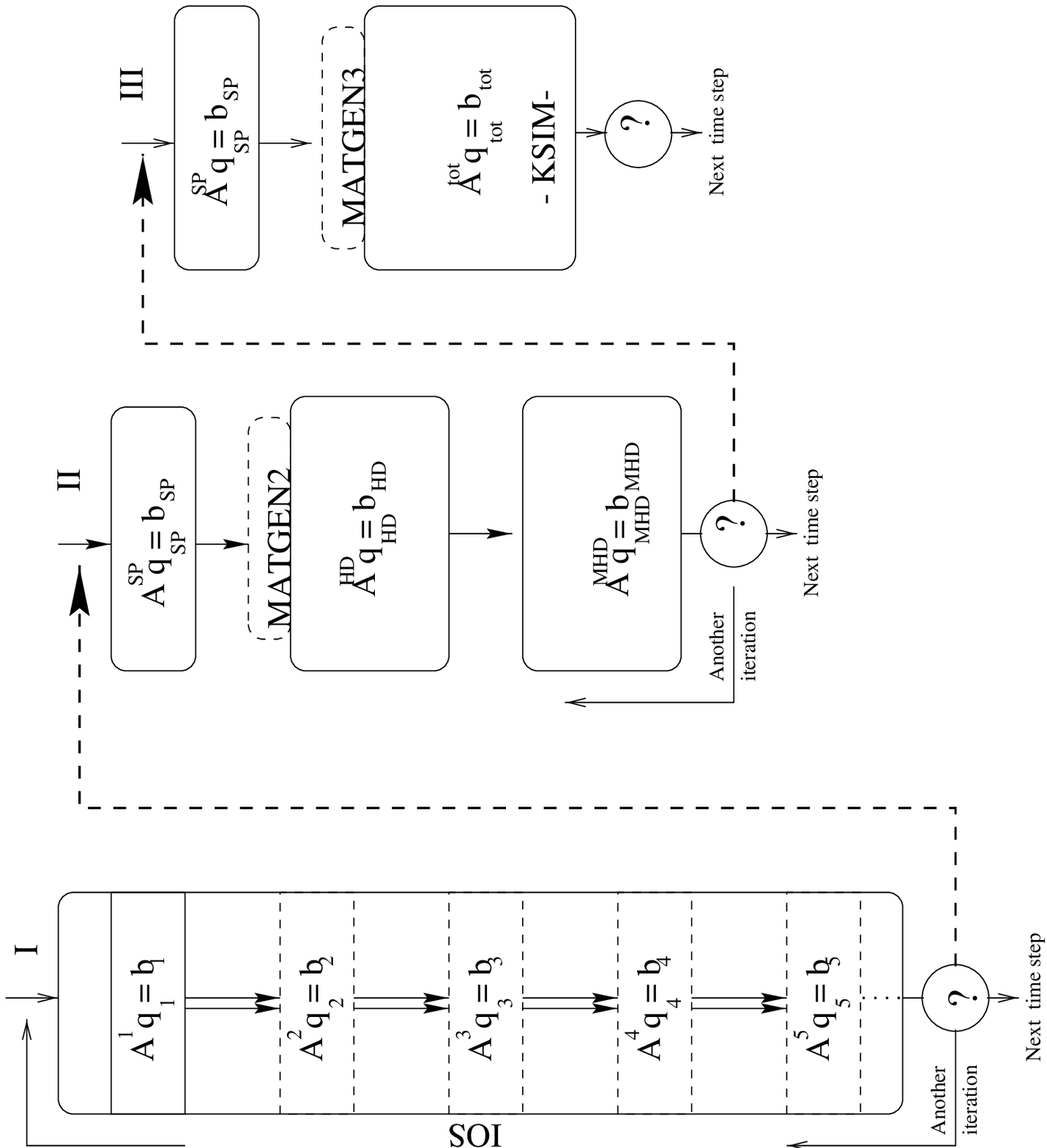}
\end{center}
\caption[ ]{\label{HSS} \small  A schematic description of the hierarchical
structure of  the algorithm. Stage I corresponds to the implicit
operator splitting approach (IOS), which is most appropriate for
following the early time-dependent phases of the flow. The solutions
obtained can then be used as initial conditions for Stage II, where
the hydro-equations are solved as a single coupled system, followed
by the  magneto component, which is again solved as a single coupled
system. Here, high spatial and temporal accuracies in combination
with the prolongation/restriction strategy may be used. Similarly,
the solutions obtained in this stage may be used as starting
solutions for Stage III, where steady solutions for the fully
coupled set of equations consisting of the zero moment of the
radiation field and the MHD equations are sought. In this stage,
pre--conditioned Krylov sub--space iterative methods are considered
to be robust and efficient. The very last stage, Stage IV,
corresponds to the case where solutions for the internal energy
equations weakly coupled with the 5D radiative transfer equation are
sought. }
\end{figure}

In an attempt to enhance both efficiency and robustness, the
hierarchical solution strategy (HSS) has been suggested
(Fig. \ref{ClusterMatrix} and \ref{HSS}, also see \citet{Hujeirat2005} and the references therein). The HSS relies
on the fact that any set of hydrodynamical, magnetohydrodynamical or
radiative equations are linearize-able and therefore can be
re-written
 in a simple matrix equation
\( A q = b,\) where $A,q,b$ are the coefficient matrix, vector of unknown variables and the vector
 of known quantities,
respectively.
Applying the defect-correction strategy \citep{Stetter1978, BoehmerStetter1984},
we may then re-write this equation as: \(A\delta q = d,\) where $\delta q$ and
$d$ denote respectively the vectors of small-time corrections and
the defect, provided  $d$ is consistent with the real mathematical
equations by construction. The matrix A in the latter equation can
be replaced by
 a variety of matrices that correspond to a sequence of numerical
 approaches that ranges from purely explicit to strongly implicit
 \citep{Hujeirat2005}. In this formulation, explicit methods arise as a
 special case, in which A is being replaced by the most easiest
 invertible-matrix: the identity matrix I. Based on this
 formulation, the Courant condition
 follows  from the requirement that the matrix A should be stable-invertible.

 Therefore,  strongly implicit and explicit methods are different
 variants of the same algebraic problem. While the former retain
 almost all the entries of the matrix A in the inversion procedure,
 the latter rely on neglecting all off-diagonal entries as well as
 crudely simplifying the diagonal elements. These methods are
 well-unified within the HSS, and that
  depending on the physical properties of the flow, a directive
  will carefully select the entries of A that are relevant for the
  problem.

In Table \ref{Table1} we have summarized properties of  several
numerical methods. Thus, as long as efficiency is concerned,
explicit methods are unrivalled candidates, provided the flows are
strongly time-dependent and compressible. {However, due to the
relatively large sound speed, these methods may stagnate both in
modeling incompressible or even weakly compressible flows. To
clarify this point, we mention that the time step-size in explicit
methods must satisfy the Courant-Friedrichs-Lewy (CFL) condition}:
\[\delta t_{exp} < \min\{\DD{\Delta x}{U + V_s}\} = \min\{\DD{\Delta x}{U(1 + 1/\cal{M})}\}, \]
where the minimum function runs over all points of the domain of
calculation. Here $\delta t_{exp}, \Delta x, U, V_s, \cal{M}$
correspond to the explicit time step-size,  space increment, velocity, sound
speed and the Mach number, respectively. Therefore, as
\({\cal{M}}\rightarrow 0 \) the flow becomes incompressible and the
time step size approaches zero; hence a stagnation of the
time-advancement procedure. We note that although in this case using
consistent implicit solution procedures is necessary, by no means it
is sufficient. Here it has been verified that standard implicit
solvers experience serious difficulties in simulating low Mach
number flows, typically found in the interior layers of stars,
planets as well as  around moving vehicles in the Earth atmosphere.

 The above discussion addresses the following questions: 1)
Relativistic fluid motions typically occur on the dynamical time
scale. The advantages of still using implicit solvers should be
clarified. 2) Multigrid methods have been shown to display
convergence which depends  weakly on the number of unknowns in the
finite space. In combination with nested iteration, the multigrid
method can solve these problems to truncation error accuracy in a
number of operations that is proportional to the number of unknowns.
   Therefore the reason for still favouring the prolongation strategy
   over multigrid or adaptive mesh refinement needs to be explained.
   3) The storage capacities of modern computers to date are capable
   of handling the entries of large matrices that correspond to the
   3D MHD equations. Thus, the reliability and credibility  of 3D
   axi-symmetric algorithms should be justified.

In fact, there are several reasons that justify using implicit
numerical procedures for modeling relativistic flows. In particular:
 \bit
\item The set of relativistic MHD equations is generically a highly coupled non-linear system,
 which gives rise to fast growing perturbations due to
non-linearities, thereby imposing a further restriction on the size
of the time step.
\item In most of the cases the horizon of a black hole represents a geometrical singularity.  The deformation
      of the geometry grows non-linearly when approaching the black hole. Thus, in order to
       capture flow-configurations  in  the vicinity of a black hole
       accurately, a non-linear distribution of the grid points is
       necessary, which, again, may destabilize explicit schemes.
\item  Depending on the evolutionary conditions, non-relativistic flows may become ultra-relativistic or vice versa.
       However, almost all non-relativistic astrophysical flows known to date are considered to be dissipative and diffusive.
       Therefore, in order to track their time-evolution reliably, the  employed numerical solver
       should be capable of treating  the corresponding second order viscous terms properly.
\item  The timescales of most astrophysical flows are considered to have a great disparity.
       Stability requirement of conditionally stable methods however
       requires that the time step size should be a small fraction
       of the shortest possible timescale. This implies that, in
       order to cover a timescale of astrophysical relevance,  an
       extremely large number of time steps would be required, which
       would give rise to prohibitive computational costs.
       Furthermore, the accumulated round off errors resulting from
       performing a large number of time-extrapolations for
       time-advancing a numerical hydrodynamical solution may easily
       cause divergence.
\item The initial conditions of most astrophysical phenomena are not known.
      Therefore, in carrying out global hydrodynamical simulations,
      the end solution should weakly depend on the initial flow
      configuration.
       Conditionally stable numerical methods, however, rely on
time-advancing of the initial conditions. \eit
 The latter reason may
explain also why using the prolongation strategy is advantageous
over classical multigrid. Worth noting  is that the main building
blocks of multigrid methods are:
 \ben
\item Restriction, i.e., down sampling of the residual errors into
coarser meshes.
\item Residual smoothing: reducing the high frequency errors by
performing several iterations, using a computationally efficient
iterative procedure such as Jacobi or Gauss-Seidel.
\item Prolongation, which relies on the interpolation from the coarse
onto finer meshes.
 \een
 The high-frequency errors here are reduced by cheap smoothing on
 the  fine meshes, whereas the low-frequency errors are reduced by
 defect correction on the coarser meshes. As the bulk of the
 algebraic operations are made on the coarser meshes, the combined
 solution procedure is considerably efficient. However, multigrid
 methods display satisfactory convergence, if the underlying flows
 are predominantly diffusion-dominated. In the case of
 advection-dominated flows, errors,  that are responsible for the
 slow convergence on the fine meshes, can be easily advected  by the
 flow on the coarser meshes, thereby reducing the coarse grid
 correction. In the case of astrophysical flows, the corresponding
 equations may change their character from Newtonian into
 ultra-relativistic or vice versa. Unlike Newtonian flows that are
 generically diffusion-dominated, relativistic flows may become
 predominantly advection-dominated, depending on how large the
 corresponding Lorentz factors are. Hence, multigrid methods may
 fail to provide the expected rate of convergence.

Finally we note that in order to model the formation and
acceleration of relativistic flows in the vicinity of ultra-compact
objects accurately, it is necessary to cover the domain of
calculation by a strongly stretched mesh. Furthermore,  Lorentz
factors enhance the  inner-coupling of the relativistic equations
and give rise to a larger spectrum of non-linearities. These
numerical difficulties combined with the need to include
sophisticated magnetic and radiative processes make the construction
of  a fully 3D algorithm, at the moment,  a computationally
unrealizable numerical task.

Therefore, in this paper, we do not intend to perform 3D global
simulations, but rather focus on the algorithmic structure of
unconditionally stable and robust 3D axi-symmetric solvers.  These
algorithms should enable us to search for stationary or
quasi-stationary solutions for the fully-coupled radiative MHD
equations in which detailed physical processes are taken into
account. \\
The paper runs as follows: in Sec. 2 we describe the relativistic
hydrodynamical equations, in Sec. 3   the transformation between the
primitive and conservative variables is described. The numerical
solution and the discretization methods are presented in Sections 4
and 5. In Sec. 6 we present the results of several test calculations
and end up with a summary in Sec. 7.


\begin{table}
\begin{minipage}{\linewidth}
\begin{tabular}{l|c|c|c}
   & Explicit & Implicit & HSS  \\\hline
\parbox[c][6ex][c]{0.15 \hsize}{solution\\ method}
   & $q^{n+1} = q^n + \delta t\,d^n$
   &  $q^{n+1} = q^n + \delta t \tilde{A}^{-1}d^*$
   &  {\scriptsize $q^{n+1}  = \alpha q^n + (1-\alpha) \delta t \tilde{A}_d^{-1} d^*$} \\ \hline
Type of flows
   & \parbox{0.23 \hsize}{Strongly time-\\dependent,\\ compressible,\\
weakly dissipative\\ HD and MHD\\ in 1, 2 and 3 dimensions}
   & \parbox{0.23 \hsize}{Stationary, \\quasi-stationary,\\
highly dissipative,\\ radiative and\\ axi-symmetric MHD-flows in 1, 2
and 3 dimensions}
   & \parbox[c][23ex][c]{0.25 \hsize}{Stationary,\\
quasi-stationary,\\ weakly compressible,\\ highly dissipative,\\ radiative
and\\ axi-symmetric MHD-flows in 1, 2 and 3 dimensions} \\ \hline
Stability
   & conditioned
   & unconditioned
   & unconditioned \\ \hline
Efficiency
   & $1$ (normalized/2D)
   & $\sim m^2$
   & $\sim m_d^2$ \\ \hline
\parbox[c][9ex][c]{0.15 \hsize}{Efficiency:\\ Enhancement\\ strategies}
   & Parallelization
   & \parbox{0.23 \hsize}{Parallelization,\\ preconditioning,\\ multigrid}
   & \parbox{0.23 \hsize}{HSS, parallelization, preconditioning, prolongation} \\ \hline
\parbox{0.15 \hsize}{Robustness:\\ Enhancement\\ strategies}
   & \parbox[c][12ex][c]{0.23 \hsize}{i. subtime-stepping \\ ii. stiff terms\\ are solved\\ semi-implicitly}
   & \parbox{0.23 \hsize}{i. multiple iteration\\ ii. reducing the time step size}
   & \parbox{0.25 \hsize}{i. multiple iteration\\ ii. reducing the time step size, HSS} \\ \hline
\parbox{0.15 \hsize}{Numerical Codes\\ Newtonian}
   & \parbox[c][12ex][c]{0.23 \hsize}{Solvers1${}^{a}$\\
      {\scriptsize
       ZEUS\&ATHENA${}^{b}$,\\
       FLASH${}^{c}$,
       NIRVANA${}^{d}$,\\
       PLUTO${}^{e}$,
       VAC${}^{f}$ }
      }
   &  Solver2${}^{g}$
   &  IRMHD${}^{h}$ \\ \hline
\parbox{0.15 \hsize}{Numerical Codes\\ Relativistic}
   & \parbox[c][15ex][c]{0.23 \hsize}{Solvers3${}^{i}$\\
      {\scriptsize
       GRMHD${}^{j}$,
       ENZO${}^{k}$,\\
       PLUTO${}^{l}$,
       HARM${}^{m}$,\\
       RAISHIN${}^{n}$,
       RAM${}^{o}$,\\
       GENESIS${}^{p}$,
       WHISKY${}^{q}$ } }
   & Solver4${}^{r}$
   & GR-I-RMHD${}^{s}$ \\ \hline
\end{tabular} \\
  \caption{\label{Table1}
  A list of only a part of the grid-oriented codes in AFD and their range of applications.
  The matrix-equations in the first row are illustrated in Sec. 4. In these equations,  $q^{n,n+1}$,
  ${\delta t}$, ${\tilde{A}}$, $\alpha$ and $d^*$ denote the vector of variables from the old and new
  time levels, time step size, a preconditioning matrix, a switch on/off parameter and a time-modified
  defect vector, respectively. ``m" in row 4 denotes the bandwidth of the corresponding matrix.}
  \smallskip
  \hrule
  \parbox{\hsize}{\footnotesize
  ${}^{a}$\citet{Bodenheimer_etal1978, Clarke1996},
  ${}^{b}$\citet{StoneNorman1992, GardinerStone2006},
  ${}^{c}$\citet{Fryxell_etal2000},
  ${}^{d}$\citet{Ziegler1998},
  ${}^{e}$\citet{MignoneBodo2003, Mignone_etal2007},
  ${}^{f}$\citet{Toth_etal1998},
  ${}^{g}$\citet{Wuchterl1990, Swesty1995},
  ${}^{h}$\citet{Hujeirat1995, Hujeirat2005, Falle2003},
  ${}^{i}$\citet{Koide_etal1999, Komissarov2004, Anninos2005,
  Meliani2007, DelZanna2007},
  ${}^{j}$\citet{DeVilliersHawley2003},
  ${}^{k}$\citet{Wang2007},
  ${}^{l}$\citet{Mignone_etal2007},
  ${}^{m}$\citet{Gammie_etal2003},
  ${}^{n}$\citet{Mizuno_etal2006},
  ${}^{o}$\citet{ZhangMacFadyen2006},
  ${}^{p}$\citet{Aloy1999},
  ${}^{q}$\citet{Baiotti_etal2003},
  ${}^{r}$\citet{Liebensdoerfer_etal2002},
  ${}^{s}$the present algorithm. }
\end{minipage}
\end{table}

\section{The hydrodynamical equations in Kerr spacetime}

In the present study we intend to numerically solve the equations of
hydrodynamics in both ultra-relativistic and purely Newtonian
regimes. Unlike the usual convention, in which the speed of light
and the gravitational constant are set to unity, we use the sound
speed as the basic measure for velocities. This is reasonable as the
radial motion of rotating flows around a compact object can be as
low as $10^{-5\pm 1}$ the speed of light, whereas it is $10^{-2\pm
1}$ of the sound speed.  Close to the event horizon, all velocities
become quantitatively comparable. This scaling enables the present
algorithm to capture the structure of slow flows accurately and
renders the appearance of terms that are extremely large or small
due to scaling effects. Additionally, the present solution procedure
is actually an extension of the purely Newtonian solver, IRMHD, into
the general relativistic regime.

\subsection{The metric}
For completeness, we develop here the equations of
hydrodynamics in the background of spacetime metric of a Kerr black
hole, using the Boyer-Lindquist coordinates ($t$, $r$, $\theta$,
$\phi$). Adopting the  3+1 split of spacetime, a line element with
the metric signature $(-,+,+,+)$ can be written as follows:
\begin{equation}
    ds^2= -\alpha^2\dt^2 + h_\mathrm{ik}(dx^i + \beta^i dt)(dx^k +
\beta^k dt), \,\, \textrm{for j,k=1, 3}.
\end{equation}
For the Kerr metric, the line element reads:
\begin{equation}
    ds^2= -(\alpha^2 -\beta_\phi\beta^\phi)\dt^2
          + 2\beta_\phi d\phi\dt + h_\mathrm{ik}dx^i dx^k ,
\end{equation}
which corresponds to a matrix of the following entries: \beq
g_{\mu\nu} = \left[ \barr{cccc}
    g_{tt} & 0 &  0 & g_{t\phi}\\
    0 & g_{rr} & 0 &  0 \\
    0 & 0 & g_{\thet\thet} & 0  \\
    g_{\phi t} & 0 & 0 & g_{\phi\phi} \earr \right]= \left[
    \barr{cccc}
    \beta_{\phi}\beta^{\phi} - \alpha^2 & 0 &  0 & \beta_{\phi}\\
    0 & h_{rr} & 0 &  0 \\
    0 & 0 & h_{\thet\thet} & 0  \\
    \beta_{\phi} & 0 & 0 & h_{\phi\phi} \earr  \right]. \eeq

The coefficients $g_{\mu\nu}$  in the Boyer-Lindquist
coordinates and their related functions are defined as follows:
  \beq \left\{ \barr{lll}
     \Delta & = & r^2 - 2r_gr+ a^2 \\
     \bar{\rho}^2& =&  r^2 + a^2 \sin^2{\theta}\\
     \Sigma^2 & = & (r^2 + a^2)^2 - a^2\Delta \cos^2{\thet} \\
     \bar{\omega} & = & \DD{\Sigma}{\bar{\rho}} \cos{\thet} \\
     \alpha & =& \DD{\bar{\rho}}{\Sigma}\sqrt{\Delta} \\
     \beta^r & =& \beta^\thet = 0, $~~~~$
     \beta^\phi =  -\DD{\omega}{c}= -2 a M_{BH} \DD{r}{c\Sigma^2} \\
     \Upsilon & = & \DD{\bar{\rho}^2 \Sigma^2}{\Delta} \cos^2{\thet} \\
     h_{rr} & =& \DD{\bar{\rho}^2}{\Delta}, $~~~~$h_{\thet\thet} = \bar{\rho}^2,$~~~~$ h_{\phi\phi}= \bar{\omega}^2 \\
    \sqrt{-g} & =& \bar{\rho}^2\cos{\thet} = \alpha
                            \sqrt{\Upsilon}.
                             \earr \right .
  \eeq
$c,\, M_{BH},\, G,\,r_g (= \DD{GM_{BH}}{c^2}),\, \alpha,\, \beta$
and ``a" denote the speed of light,
  mass of the black hole, the gravitational constant, the
  gravitational radius, the lapse and shift functions and the
  Kerr-spin parameter, respectively. In writing these expressions,
  we made use of the coordinate transformation $\bar{\thet} = \pi/2
  - \thet$, where we use the latitude $\thet$ instead of the polar distance
  angle $\bar{\thet}$; hence the appearance of "$\cos$" instead of "$\sin$" in
  the metric terms.

\subsection{The governing equations}
Following the internal energy formulation of
\citet{Wilson1972} and \citet{Hawley_etal1984a, Hawley_etal1984b},
we derive the hydrodynamical equations from the four-velocity $u^\mu
u_\mu = - c^2$, the conservation of baryonic number $\nabla_\mu
(\rho u^\mu) = 0,$ the parallel component of the stress-energy
conservation equation $u_\nu\nabla_\mu T^{\mu\nu}=0$ (to derive the
internal energy equation) and from the transverse components
$(g_{\xi\nu}+ u_\xi u_\nu) \nabla_\mu T^{\mu\nu}=0$ (to derive the
momentum equations). \\
For viscous flows, the stress energy tensor
reads:
\beq T^{\mu\nu}= \mm{{T_\mm{PF}^{\mu\nu}} +
\left\{\underline{T_\mm{Vis}^{\mu\nu}}\right\}} = \rho~h~u^\mu u^\nu
+ P~g^{\mu\nu} + \left\{
 \underline{
-\eta [ \bar{\sigma}^{\mu\nu} + \DD{\Theta}{3}h^{\mu\nu}]}
 \right\},
 \label{SET}\eeq
where $\mm{{T_\mm{PF}^{\mu\nu}}, ~T_\mm{Vis}^{\mu\nu}}$ denote the
stress energy tensor due to perfect and viscous flows, respectively.
{\rm P}, $\eta,~\Theta,~$ are the pressure, which is calculated from
the equation of state corresponding to polytropic or to an ideal
gas, the dynamical viscosity which is assumed to be identical to the
shear viscosity,  and $\Theta~(\doteq\nabla_\mu u ^\mu)$ which
measures the divergence or convergence of the fluid world lines,
respectively. $h^{\mu\nu}= u^\mu u^\nu + g^{\mu\nu}$ is the spatial
projection tensor, whereas $\bar{\sigma}$ corresponds to the
symmetric spatial shear tensor: \( \bar{\sigma}^{\mu\nu} =
\nabla_\varsigma u^\mu h^{\varsigma\nu} + \nabla_\varsigma u^\nu
h^{\varsigma\mu} .\) In the case of an ideal gas, the pressure and
enthalpy read:
 \beq
 \barr{lll}
 {\rm P} & = &(\gamma-1)C_\mm{V}\rho T = (\gamma-1)\rho {\mathcal{E}}\\
 h & = & c^2 + {\mathcal{E}} + {P}/{\rho} = c^2 + \gamma {\mathcal{E}},
\earr \eeq where $\gamma,~C_\mm{V}$ and ${\mathcal{E}}$ denote the
adiabatic index, specific heat and  internal energy   measured in
the local rest frame of the fluid. { For clarity, we re-write the
hydrodynamical equations in flux conservative form}:
 \beq \rm \DD{\D D}{\D  t} +
\DD{1}{\sqrt{-g}} \DD{\D}{\D r} (\sqrt{-g}D V^r)
                 + \DD{1}{\sqrt{-g}} \DD{\D}{\D \thet} (\sqrt{-g}D V^\thet) = 0
\eeq

\[
\rm
\DD{\D M_r}{\D  t} + \DD{1}{\sqrt{-g}} \DD{\D}{\D r} (\sqrt{-g}M_r V^r)
                  + \DD{1}{\sqrt{-g}} \DD{\D}{\D \thet}(\sqrt{-g} M_r V^\thet)
= \]
\beq \hspace*{0.5 \hsize}
{}- \DD{\D P}{\D r} - \DD{1}{2}(\DD{M^\mu
  M^\nu}{M^t})\DD{\D g_{\mu\nu}}{\D r}  + L2^{r}_{vis}
\eeq

\[
\rm
\DD{\D M_\thet}{\D  t} + \DD{1}{\sqrt{-g}} \DD{\D}{\D r}(\sqrt{-g}M_\thet V^r)
                       + \DD{1}{\sqrt{-g}} \DD{\D}{\D \thet}(\sqrt{-g} M_\thet V^\thet)
= \]
\beq \hspace*{0.5 \hsize}
{}- \DD{\D P}{\D \thet} -
  \DD{1}{2}(\DD{M^\mu M^\nu}{M^t})\DD{\D g_{\mu\nu}}{\D \thet} + L2^{\thet}_{vis}
\eeq

\beq
\rm
\DD{\D M_\phi}{\D  t} + \DD{1}{\sqrt{-g}} \DD{\D}{\D r} (\sqrt{-g}M_\phi V^r)
                      + \DD{1}{\sqrt{-g}} \DD{\D}{\D \thet} (\sqrt{-g} M_\phi V^\thet) = L2^{\phi}_{vis}
\eeq

\[
\rm
\DD{\D {\cal{E}}^d}{\D  t} + \DD{1}{\sqrt{-g}} \DD{\D}{\D r} (\sqrt{-g} {\cal{E}}^d V^r)
                           + \DD{1}{\sqrt{-g}} \DD{\D}{\D \thet} (\sqrt{-g}  {\cal{E}}^d V^\thet)
= \]
\[
{}-(\gamma-1)\,{\cal{E}}^d \,[\DD{\D u^t}{\D  t}
                 + \DD{1}{\sqrt{-g}} \DD{\D}{\D r} (\sqrt{-g} u^t V^r)
                 + \DD{1}{\sqrt{-g}} \DD{\D}{\D \thet} (\sqrt{-g} u^t
                 V^\thet)\]
\beq
                 + \Phi + \Gamma - \Lambda,
                 \label{Equi5}
\eeq

where $\mm{D} = \rho u^t$ is the modified relativistic mass density.
$\rm{M}_\mu$ are the four-momenta: \(\rm (M_t,M_r,M_\thet,M_\phi)
\doteq {\DroI}(u_t,u_r,u_\thet,u_\phi),\) where \( {\DroI} \doteq D
h,\) and $u^t$ is the time-like velocity, \( V^\mu= u^\mu/u^t\) is
the transport velocity. \(L2^{\xi}_{vis}\) are the spatial
projections of the viscous stress energy tensor
$\mm{T^{\mu\nu}_{Vis}}~$ (see Eq. \ref{SET}) in the respective
direction. These are obtained from the projection of the viscous
tensor along the vector normal to the hyperspace, i.e., constant in
time:
\[
L2^{\xi}_{vis} = \nabla_\mu T_\mm{Vis}^{\mu\xi} = \bar{\partial}_\mu
T_\mm{Vis}^{\mu\xi} + \Gamma_{\mu\lambda}^{\xi}
T_\mm{Vis}^{\mu\lambda},
\]
where $\xi=\{r,\theta,\varphi\}$. $\nabla_\mu$ corresponds to the
spatial divergence of a tensor taken in the Boyer-Lindquist
coordinates and $\Gamma_{\mu\lambda}^{\xi}$ are the Christoffel's
symbols of the
second kind.\\
 From the collection of the numerous viscous terms, we
only consider the dominant second order operators, that are set to
degenerate into the classical non-relativistic formulation of the
Navier-Stokes equations if the sound speed becomes smaller than a critical value
\citep{Tassoul1978}.\\
The viscosity coefficient here is based on the alpha-turbulent
description, $\alpha_t,$ modified to respect causality. Hence the
dynamical viscosity reads:
 \beq \eta = \rho <l><\mbox{v}_{turb}> =  \rho <l>  [\alpha V_s/u^t] =
 \DD{D}{(u^t)^2}   [\alpha <l> V_s] = \DD{D}{(u^t)^2} \nu_t, \eeq
where $<>$ denotes a turbulent mean, $\mbox{v}_{turb}$ is the
relativistic turbulent velocity and $\nu_t$ the relativistic
turbulent velocity coefficient and $\alpha_t$ is a constant of order
unity.\\

Equation (\ref{Equi5}) describes the time-evolution of the
relativistically modified internal energy ${\cal{E}}^\mm{d} =
\mm{~D\,C_V~T,}$ where T is the plasma temperature. $\Phi,~\Gamma,~
\Lambda$  correspond to heat function due to turbulent dissipation,
other heating and cooling functions, respectively. Using the
transformation \(\rm q_{\mu}= g_{\mu\nu} q^{\nu}\), we may define
the transport velocities $\rm{V}^\mu$ as follows
\citep[see][]{Hawley_etal1984a, Hawley_etal1984b}: \beq \rm
   \left\{
  \barr{lllll}
  \rm
     u_r  & = & g_{rr}~{u^r} & = & g_{rr} ~ u^t ~({V^r}/c)\\
     u_\thet  & = & g_{\thet\thet} ~{u^\thet} & = & g_{\thet\thet}~ u^t~({V^\thet}/c)\\
     u_t  & = & g_{tt} ~u^t + g_{t \phi}~u^\phi & = & u^t~[g_{tt} + g_{t \phi}~(V^\phi/c)] \\
     u_\phi  & = & g_{t\phi} ~u^t + g_{\phi \phi}~u^\phi & = & u^t~[g_{t\phi} + g_{\phi \phi}~(V^\phi/c)] .\\
  \earr  \right .
 \eeq
 The corresponding relativistic 4-momenta then read:

 \beq
 \rm
   \left\{
  \barr{lll}
     M^t     & = & \DroI u^t \doteq   \DroII  \\
     M^r     & = & \DroI   [{V^r}/{c}] \\
     M^\thet & = & \DroI   [{V^\thet}/{c}] \\
     M^\phi  & = & \DroI   [{V^\phi}/{c}], \\
  \earr  \right .
  \eeq
  from which the covariant 4-momenta  may be  obtained:
\beq
\rm
   \left\{
  \barr{lll}
     M_t       & = & g_{tt}~ M^t + g_{t\phi} ~M^\phi \\
     M_r       & = & g_{rr} ~M^r  \\
     M_\thet   & = & g_{\thet\thet} ~M^\thet \\
     M_\phi    & = & g_{\phi\phi}~M^\phi + g_{\phi t}~ M^t.
  \earr  \right .
  \eeq

We note that by using this formulation of the HD-equations in
combination with finite volume discretization, mass and momenta  are
conserved up to small discretization errors.  This is necessary in
order to assure that inflowing  non-rotating matter would not gain
angular momenta though
it must rotate in the ergosphere. 

 \subsection{Non-dimensional formulation}
  The algorithm presented here should be capable of modeling the
  time-evolution of hydrodynamical flows both in the
  non-relativistic as well as in the extreme-relativistic regimes.
  In order to avoid the appearing of extremely small coefficients in
  the equations, the scaling variables listed in Table (\ref{Tablescaling}) are adopted.

\begin{table}[Hhtb]
\begin{tabular}{llll }
Scaling variables &                     & \hspace*{1.0cm}& Example
(supermassive BH)
\\\hline
Mass:             &   $\tilde{\mathcal M}$ & &$3\times 10^8M_{\odot}$    \\
Accretion rate:   &   $\hspace{1.45ex}\tilde{\rule[0.0ex]{0ex}{2ex}}\hspace{-1.9ex}\Mdot$ &&  $10^{-2}\Mdot_{Edd}$  \\
Distance:         &   $\tilde{R}$  && $3R_\mm{S}$ \\
Temperature:      &   $\tilde{\mathcal T}$    &&  $10^8 ~\mm{K}$  \\
Velocities:       &   $\tilde{V}$ &&  $\tilde{V_\mm{S}}$ \\
Density:          &   $\tilde{\rho}$ && $10^{-12}~~ \mm{g\,cm^{-3}} $  \\
 \hline
 \end{tabular}
\caption{Scaling variables that might be used to re-write the
equations in non-dimensional form. In this table 
$R_S~~ (= 2 r_g)$ stands for the Schwarzschild radius 
($r_g $ is the gravitational radius)  
and $\tilde{V_\mm{S}}$ is the sound speed.}
\label{Tablescaling}
\end{table}

We now introduce the following additional non-dimensional
parameters: \beq \rm
   \left\{ \barr{lllllllllllllll}
     \eps_{BH} &=& {r_g}/{\tilde{R}}, ~~~~\eps_{1} &=&
  \DD{\tilde{V_\mm{S}}}{\tilde{V_\phi}}, ~~~~ \eps_{6} &=&
  \DD{\tilde{V_\mm{S}}}{c}, ~~~~\bar{a} &=& a ({r_g}/{\tilde{R}})
  \earr  \right\} . \eeq
where a is the black hole spin.

The normalization of the 4-velocity and momentum yields:
{\footnotesize
\beq
 (U^t)^2 \left[g_{tt} + 2 \left(\DD{\eps_6}{\eps_1}\right)^2 g_{t\phi} V^\phi
                 + {\eps_6}^2 g_{rr} (V^r)^2
         +   {\eps_6}^2 g_{\thet\thet} (V^\thet)^2
               + \left(\DD{\eps_6}{\eps_1}\right)^2g_{\phi\phi}(V^\phi)^2\right] = -1,
\eeq
}
and
{\small
\beq
 g^{tt} M^2_t + 2 \left(\DD{\eps_6}{\eps_1}\right)^3 g^{t\phi}M_t M_\phi
              + {\eps_6}^2 g^{rr}M^2_r + {\eps_6}^2 g^{\thet\thet}M^2_\thet
          + \left(\DD{\eps_6}{\eps_1}\right)^2 g^{\phi\phi}M^2_\phi  = -\DroI^2,
\eeq
}
where
{\small
\beq
\rm
   \left\{
  \barr{lll}
     g^{tt}          &=&  -1/\alpha^2\\
     g^{t\phi}       &=&   a\,r/(\alpha\Sigma)^2 \\
     g^{\phi\phi}    &=&  [1- (\DD{4}{\alpha^2})(\DD{a\,r}{\Sigma^2})]\\
     g^{rr}          &=&  1/g_{rr} = \Delta/{\bar{\rho}}^2\\
     g^{\thet\thet}  &=&  1/g_{\thet\thet} = 1/{\bar{\rho}}^2.
  \earr  \right .
  \eeq
}
We may write the   equations of hydrodynamics in non-dimensional
formulation
in a manner that they smoothly adopt the Newtonian form in the non-relativistic
regime:
\beq \rm \DD{\D D}{\D  t} + \DD{1}{\sqrt{-g}} \DD{\D}{\D r}\left(\sqrt{\DD{-g}{g_{rr}}}\,\, D U\right)
                          + \DD{1}{\sqrt{-g}} \DD{\D}{\D \thet}\left(\sqrt{\DD{-g}{g_{\thet\thet}}} \,\,D V\right)
              = 0
\eeq
\[
\rm
\DD{\D M_r}{\D t} + \DD{1}{\sqrt{-g}} \DD{\D}{\D r}\left(\sqrt{\DD{-g}{g_{rr}}} \,\,M_r U\right)
                  + \DD{1}{\sqrt{-g}} \DD{\D}{\D \thet}\left(\sqrt{\DD{-g}{g_{\thet\thet}}} \,\,M_r V\right)
          =
\]
\beq \rm \hspace*{0.45 \hsize}
{}- \DD{\D P}{\D r} - \DD{1}{2}\left(\DD{M^\mu M^\nu}{M^t}\right)\DD{\D g_{\mu\nu}}{\D r} + Q^{r}_{vis}
\eeq
\[
\rm
\DD{\D M_\thet}{\D  t} + \DD{1}{\sqrt{-g}} \DD{\D}{\D r} \left(\sqrt{\DD{-g}{g_{rr}}} \,\,M_{\thet} U\right)
                       + \DD{1}{\sqrt{-g}} \DD{\D}{\D \thet} \left(\sqrt{\DD{-g}{g_{\thet\thet}}} \,\,M_{\thet} V\right)
               =
\]
\beq \rm \hspace*{0.45 \hsize}
{}- \DD{\D P}{\D \thet}
  - \DD{1}{2}\left(\DD{M^\mu M^\nu}{M^t}\right)\DD{\D g_{\mu\nu}}{\D \thet} + Q^{\thet}_{vis}
\eeq
\beq
\rm
\DD{\D M_\phi}{\D  t} + \DD{1}{\sqrt{-g}} \DD{\D}{\D r} \left(\sqrt{\DD{-g}{g_{rr}}} \,\,M_{\phi} U\right)
                      + \DD{1}{\sqrt{-g}} \DD{\D}{\D \thet} \left(\sqrt{\DD{g}{g_{\thet\thet}}} \,\,M_{\phi} V\right)
=  Q^{\phi}_{vis}
\eeq
\[
\rm
\DD{\D {\cal{E}}^d }{\D  t} + \DD{1}{\sqrt{-g}} \DD{\D}{\D r} \left(\sqrt{\DD{-g}{g_{rr}}} \,\,{\cal{E}}^d U\right)
   + \DD{1}{\sqrt{-g}} \DD{\D}{\D \thet} (\sqrt{\DD{-g}{g_{\thet\thet}}}\,\, {\cal{E}}^d V) =
\]
\[
{\rm \hspace*{0.02 \hsize}
{}-(\gamma-1)\,{\cal{E}}^d \,\left[\DD{\D u^t}{\D  t}
  + \DD{1}{\sqrt{-g}} \DD{\D}{\D r} \left(\sqrt{\DD{-g}{g_{rr}}}\,\, u^t U\right)
  + \DD{1}{\sqrt{-g}} \DD{\D}{\D \thet} \left(\sqrt{\DD{-g}{g_{\thet\thet}}}\,\, u^t V\right)
  \right]} \]
\beq
  + \Phi + \Gamma - \Lambda,   \label{eq:E_nondim} \eeq
where  \({\rm U = V^r\,\sqrt{g_{rr}}, V =
V^\thet\,\sqrt{g_{\thet\thet}},
   V^\phi = \bar{V}^\phi\,\sqrt{g_{\phi\phi}},\,  h= 1 +
 (\DD{\eps_6^2}{\gamma-1})\,T.}\) \\
 For flows approaching rotating
 black holes,  the angular momentum is defined as: \(M_\phi = \DroII
 \sqrt{g_{\phi\phi}} \, (V^\phi + V_{FDE}^\phi), \) where ${V_{FDE}}$
 denotes the rotation of the flow that is induced due to the
 frame-dragging effect:
 ${V^\phi_{FDE}} = {\beta^\phi}/{\sqrt{g_{\phi\phi}}}.$ \\ 
Note that the radial velocity in this formulation approaches the
speed of light as the matter crosses the event horizon.
\section{The primitive variables }

 The above set of equations describes the time-evolution of the
 conserved quantities \(D, M_i\) and ${\cal{E}}^d.$ However, the
 equation of state, the rate of transport, the applied work, cooling
 and heating rates are functions of essentially the primitive
 variables
 \(\rho,  V^i\) and $T$. \\
 Since the relation between the primitive and conservative variables
is rather non-linear, an iterative solution procedure should be
employed.

 We note that the 4-momenta must satisfy the normalization condition:
  \(M_\mu M^\mu = -M^2 \doteq -\DroI^2 \). This is equivalent to
  solve the following equation for $M_t$:
  {\small
  \beq M_\mu M^\mu =
  g^{\mu\nu}M_\mu M_\nu =g^{tt} M^2_t + 2 g^{t\phi}M_t M_\phi
              + g^{rr}M^2_r + g^{\thet\thet}M^2_\thet + g^{\phi\phi}M^2_\phi = -\DroI^2.
  \eeq
  }

  Taking into account that the quantities $M_r,\,M_\thet,\,M_\phi$
  are known at the end of each time step, we may substitute them in
  Eq. (\ref{eq:E_nondim}) to obtain a quadratic equation for $M_t$, i.e., \beq
     \aleph M^2_t + \beth  M_t + \gimel = 0, \eeq where \(\aleph=
  g^{tt},\, \beth= 2 g^{t\phi}M_\phi \textrm{\,\,and\,\,} \gimel =
  g^{rr}M^2_r + g^{\thet\thet}M^2_\thet + g^{\phi\phi}M^2_\phi.\)\\
  Having obtained $M_t$, the contravariant quantity
  $M^t$ can be computed using the transformation:
  \(M^t = g^{tt}M_t + g^{t\phi}M_\phi,\) whereas the global Lorentz factor is obtained from:
  \(u^t = M^t/D h.\)
  Using Equation (\ref{eq:E_nondim}), the pressure can be
  calculated then from the relation:
  $P = (\gamma-1)\mathcal{E}^\mm{d}/u^\mm{t}.$



\begin{figure}[htb]
\centering {
\includegraphics*[width=\hsize, bb=5 160 587 835,clip]{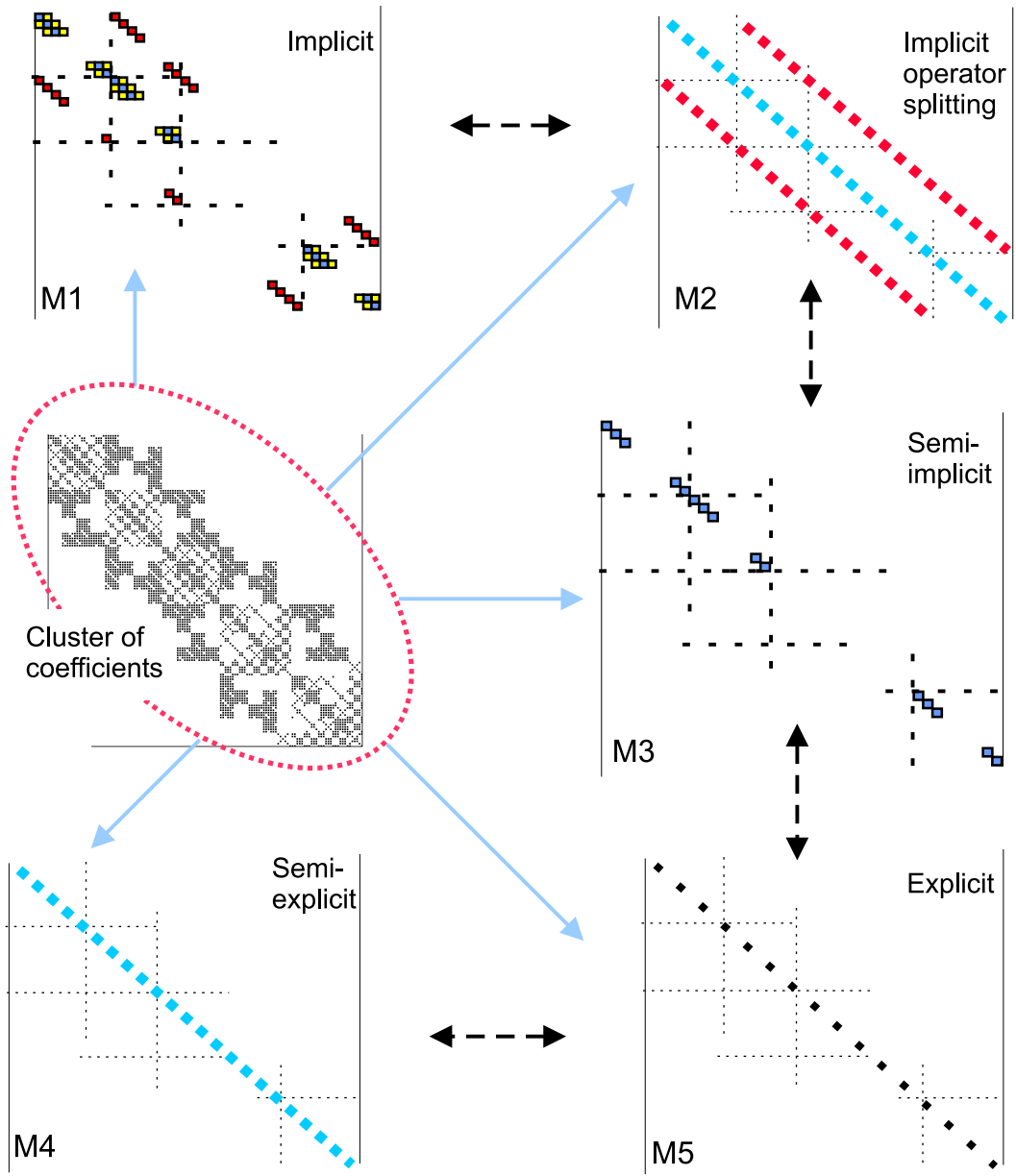}
}
\caption{\label{ClusterMatrix} \small A schematic description of the hierarchical
clustering of the coefficient matrix employed by the algorithm.
Here, a cluster of coefficients is computed in the very early stage.
The matrix-generator then selects the entries to be used for
constructing the matrix coefficient appropriate for the solution
procedure.  Depending on the matrix used, the solution method may
range from purely explicit to fully implicit. Interchange between
solution methods is possible, as modifying, adding or removing
entries is directly maintainable. }
\end{figure}
%

\section{The hierarchical solution strategy - HSS  }
The set of hydrodynamical equations are solved within the context of
the hierarchical solution strategy (HSS, see \citet{Hujeirat2005}). HSS is
based on constructing a  coefficient matrix $A$, which results from
linearizing the complete set of equations in a fully implicit
manner. Noting that the conservative formulation of the HD-equations
yields a matrix coefficient that is highly sparse,  it is reasonable
to design a procedure which selects the entries for constructing the
approximate matrix $\hat{A}$ most appropriate for the flow problem.
Depending on the structure of  $\hat{A}$, a suitable iterative
method within the context of defect-correction method may be
employed to assure consistency and convergence.\\

For example, if we want to simulate a two-dimensional weakly compressible, non-magnetized and
 non-radiating flow between two concentric spheres, then  the
above-mentioned procedure is set to select the entries from the
cluster of coefficients that corresponding just to the  equations to
be solved (see Fig. \ref{ClusterMatrix}), which are used then to construct the
preconditioning $\hat{A}$.

To clarify the procedure, we re-write the set of equations in a conservative vector form:

\beq
 \DD{\partial \vec{q}}{\partial t} + L_\mm{r,rr} \vec{F} +
L_\mm{\theta,\theta\theta} \vec{G}  = \vec{f} , \label{eq:conservative_vector_form} \eeq where $\vec{F}$
and $\vec{G}$  are fluxes of $q$, and $ L_\mm{r,rr},\,
L_\mm{\theta,\theta\theta}$ are first and second order operators
 that describe the advection and diffusion of  the vector variables
$\vec{q}$ in $\mm{r}$ and $\theta$  directions. $\vec{f}$
corresponds
to the vector of source functions.\\

{ In order to enhance the mathematical consistency and increase the
spatial and temporal accuracies of the numerical scheme without a
substantial increase of the computational costs,
 we adopt the so-called prediction-correction iteration procedure.
Therefore, we re-write Eq. (\ref{eq:conservative_vector_form})
  in residual form:
$R =\DD{\partial \vec{q}}{\partial t} + L_\mm{r,rr} \vec{F} +   L_\mm{\theta,\theta\theta} \vec{G}  - \vec{f} = 0 $,
 and  adopting a five star staggered grid discretization, we may
apply the Newton-linearization  to calculate the  Jacobian,
$J_\mm{m1,n1}\doteq \DD{\partial R_\mm{m1}}{\partial q_\mm{n1}}$, where $\mm{m1,n1}$ are integers that
run over the number
of equations and variables.
The solution can be obtained then as follows:
\[
\vec{q}^{i+1} = \vec{q}^{i} - J^{-1}_\mm{m1,n1}R^{i},
\]
where $i$ is the iteration level. By inspection of the Jacobian J,
it can be easily verified that it  has the following block matrix
structure: }
\[
\DD{{\delta q}_\mm{j,k}}{\delta t} + \underline{S}^\mm{r}{\delta
q}_\mm{j-1,k} + {D}^\mm{r}{\delta q}_\mm{j,k}
                + \overline{S}^\mm{r}{\delta q}_\mm{j+1,k} \]
\beq
 + \underline{S}^\mm{\theta}{\delta q}_\mm{j,k-1} + {D}^\mm{\theta}{\delta q}_\mm{j,k}
                + \overline{S}^\mm{\theta}{\delta q}_\mm{j,k+1}
= RHS^\mm{n}_\mm{j,k}, \label{eq:block-matrix-structure}
\eeq
{ where ${\delta q} = q^{i+1} - q^{i},$ and which, in the linear case, reduces
to time-difference  of $q$.}
The subscripts ``j'' and ``k'' denote
the grid-numbering in the $\mm{r}$ and $\theta$ directions,
respectively, and $RHS^\mm{n}=[\vec{f}- L_\mm{r,rr} \vec{F} -
L_\mm{\theta,\theta\theta} \vec{G}]^\mm{n}$.
 $\underline{S}^\mm{r,\theta}$ and $\overline{S}^\mm{r,\theta}$
 mark the sub-diagonal and super-diagonal block matrices, respectively.
 ${D}^\mm{r,\theta}$ corresponds to diagonal block matrices.\\
To outline the directional dependence of the block matrices, we
re-write Eq. (\ref{eq:conservative_vector_form}) in a more compact form:
 \beq
 \begin{array}{lll}
 & {\hspace*{1.6ex}}\overline{S}^\mm{\theta}{\delta q}_\mm{j,k+1} & \\
+ \underline{S}^\mm{r}{\delta q}_\mm{j-1,k} & +
{D}_\mm{mod}{\delta q}_\mm{j,k} &
  + \overline{S}^\mm{r}{\delta q}_\mm{j+1,k}   = RHS^\mm{n}_\mm{j,k} \\
& + \underline{S}^\mm{\theta}{\delta q}_\mm{j,k-1}, &
\end{array}
\eeq
  where \({D}_\mm{mod} = I/{\delta t} +
{D}^\mm{r}+ {D}^\mm{\theta}.\) \\
 This equation  gives rise to at least four
different types of solution procedures: \ben
\item Classical explicit methods are very special cases in which
the sub- and super-diagonal block matrices together with
         ${D}^\mm{r}$ and ${D}^\mm{\theta}$ are neglected. The only
    matrix to be retained here is $ (1/{\delta t})\,\times\,$(the
    identity matrix), i.e., the first term on the LHS of Eq. (\ref{eq:block-matrix-structure}).
    This yields the vector equation (see M5/Fig. \ref{ClusterMatrix}): \beq
        [\DD{I}{\delta t}]{\delta q}_\mm{j,k}  = RHS^\mm{n}_\mm{j,k}.
    \eeq
\item Semi-explicit methods are obtained by preserving the diagonal
entries,
      $d_\mm{j,k},$ of the block diagonal matrix ${D}_\mm{mod}$ (see
      M4/Fig. \ref{ClusterMatrix}). This method has been verified to be numerically
      stable even when large Courant-Friedrichs-Levy (CFL) numbers
      are used. In particular, this method is absolutely stable if
      the flow is viscous-dominated.
\item Semi-implicit methods are recovered when neglecting the sub- and super-diagonal
  block matrices only, but retaining the block diagonal matrices
(see M3/Fig. \ref{ClusterMatrix}). In this case
      the matrix equation reads:
    \beq
       {D}_\mm{mod}{\delta q}_\mm{j,k} = RHS^\mm{n}_\mm{j,k}.
    \eeq
       We note that inverting ${D}_\mm{mod}$ is a straightforward procedure, which can be maintained
       analytically or numerically.
\item A fully implicit solution procedure requires retaining
       all the  block matrices on the LHS of Eq. (\ref{eq:block-matrix-structure}). This yields a
     global matrix that is highly sparse (M1/Fig. \ref{ClusterMatrix}).  In this case,
     semi-direct methods such as the ``Approximate Factorization
     Method''
     \citep[AFM: ][]{BeamWarming1978}
     and the ``Line Gauss-Seidel Relaxation Method''
     \citep[-LGS: ][]{MacCormack1989}
       are considered to be efficient preconditioners { for solving the set of radiative
        MHD-equations within the context of defect-correction
iteration method (see \citet{Hujeirat2005} and the references therein).
Furthermore, Krylov sub-iterative methods may prove
      to be more efficient and robust than the above-mentioned semi-direct methods.}
\een

In the case that only stationary solutions are sought, convergence
to steady state can be accelerated by adopting the so-called
``Residual Smoothing Method''
\citep{Hujeirat2005}
This method is based on associating a time step size with the local CFL-number at
each grid point.  While this strategy is efficient
    at providing quasi-stationary solutions within a reasonable number of iterations,  it
is incapable at providing
physically meaningful time scales for features that possess
quasi-stationary behaviour. Here we suggest to use the obtained
quasi-stationary solutions as initial configuration and re-start
the calculations using a uniform and physically relevant time step. 

\section{Numerical techniques}
In this section we briefly describe several algorithmic steps for
solving the continuity equation and the generalization procedure.
\begin{figure}[htb]
\centering {
\includegraphics*[width=0.435 \hsize, bb=65 290 480 630,clip]{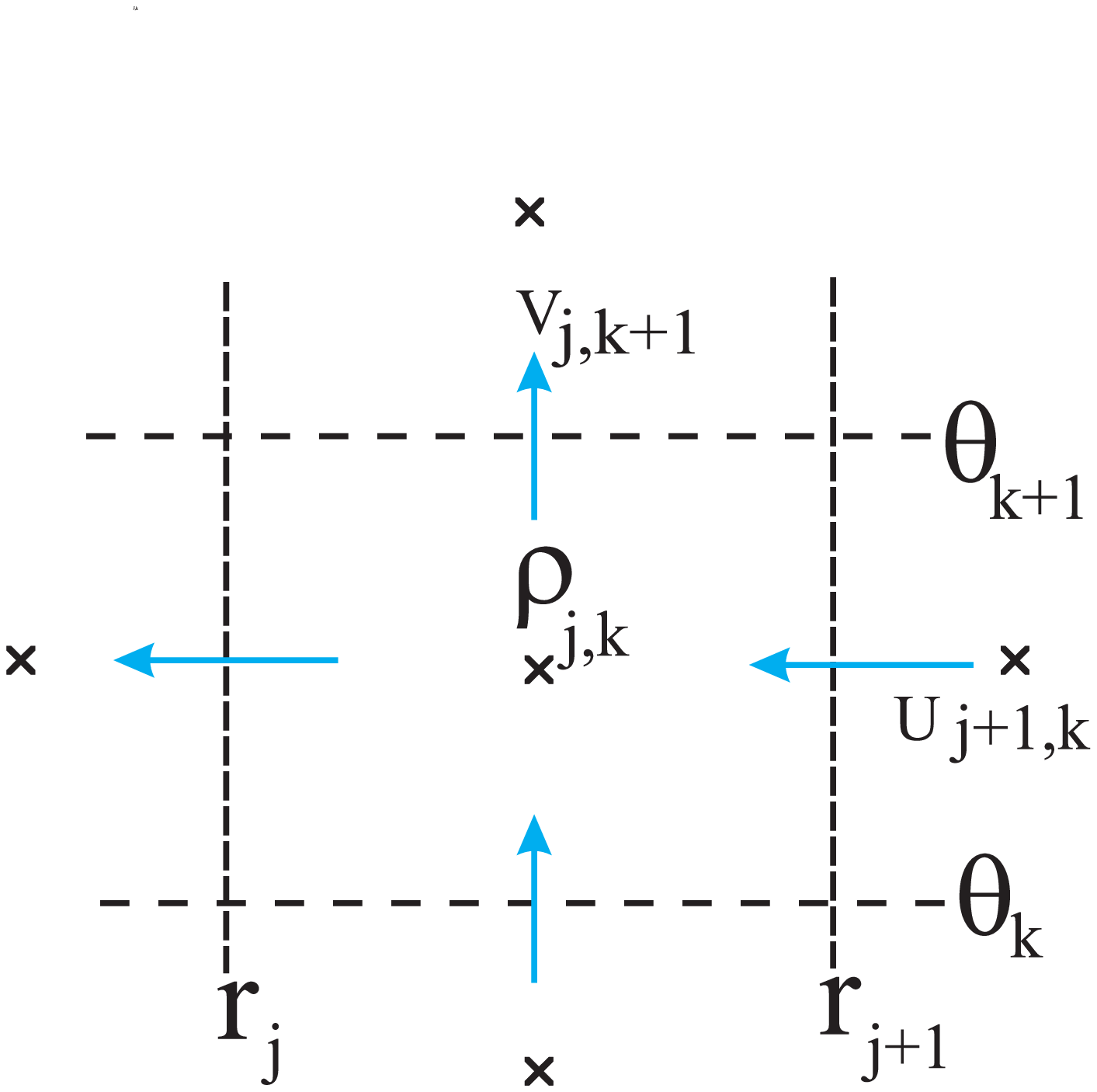}
}
\caption{\small The location of the variables  using the staggered
grid discretization. } \label{Staggered}
\end{figure}
\ben
\item The continuity equation is discretized using the staggered grid
strategy within the context of finite volume philosophy (Fig. \ref{Staggered}). 

{\footnotesize
\[
 \DD{1}{\sqrt{-g}} \DD{\D}{\D r} \left.\left(\sqrt{-g/g_{rr}} D U\right)\right|_\mathrm{j,k}
\]
\beq
\Rightarrow \left.\DD{[\sqrt{{-g}/{g_{rr}}}\, U
\overrightarrow{D^r}]^{r_{j+1}}_{r_j}}{[\int{\sqrt{-g}
dr}]^{r_{j+1}}_{r_j}}\right|_\mathrm{k} =
 \left.\DD{[\sqrt{{-g}/{g_{rr}}}\, U
\overrightarrow{D^r}]_{r=r_{j+1}} - [\sqrt{{-g}/{g_{rr}}}\, U
\overrightarrow{D^r}]_{r=r_{j}}}
 {[\int{\sqrt{-g}dr}]^{r_{j+1}}_{r_j}}\right|_\mathrm{k},
\eeq
\[
  \DD{1}{\sqrt{-g}} \DD{\D}{\D \thet} \left.\left(\sqrt{-g/g_{\thet\thet}}\right) D V\right|_\mathrm{j,k}
\]
\beq
\Rightarrow \left.\DD{[\sqrt{-g/g_{\thet\thet}} V
\overrightarrow{D^\theta}]^{\theta_{k+1}}_{\theta_k}}{[\int{\sqrt{-g/g_{\thet\thet}}
d \theta}]^{\theta_{k+1}}_{\theta_k}}\right|_\mathrm{j} =
\left.\DD{[\sqrt{-g/g_{\thet\thet}} V \overrightarrow{D^\theta}]_{\theta_{k+1}}-
[\sqrt{-g/g_{\thet\thet}} V \overrightarrow{D^\theta}]_{\theta_{k}}}
{[\int{\sqrt{-g} d \theta}]^{\theta_{k+1}}_{\theta_k}}\right|_\mathrm{j},
\eeq
}
{\footnotesize
where
\beq
\overrightarrow{D^r}_{j,k}= \left\{ \begin{array}{ll}
                          D_{j,k} + f^r_{j,k} & \textrm{if}\,\, U_{j,k} < 0  \\
                          D_{j-1,k} + f^r_{j-1,k}& \textrm{if}\,\, U_{j,k}
\ge 0
                        \end{array} \right.
            \, ; \,
\overrightarrow{D^\theta}_{j,k}= \left\{ \begin{array}{ll}
                        D_{j,k} + f^\theta_{j,k} &  \textrm{if}\,\, V_{j,k} < 0  \\
                        D_{j,k-1} + f^\theta_{j,k-1}& \textrm{if}\,\, V_{j,k}
\ge 0
                      \end{array} \right.
\eeq
}
The functions $f^r\textrm{ and }f^\theta$ are corrections for
maintaining higher order spatial accuracies.
\item In order to obtain second order temporal accuracy, we write the  continuity
equation as follows:
 \beq \DD{\delta D}{\delta  t} + \vartheta L1^{n+1}(D) +
(1-\vartheta) L1^{n}(D) = 0, \eeq
 where $\vartheta$ denotes the parameter of the stabilized
Crank-Nicolson method for achieving second order temporal accuracy.
$L1_h$ resembles the advection operator at the new time level (n+1)
and the old time level (n) and $\delta D = D^{n+1} - D^{n}. $
Taylor-expanding the variable $D^{n+1}$ in time and considering
first order terms only, the continuity equation gets the following
form: \beq \DD{\delta D}{\delta  t} + \vartheta L1^{n+1}(\delta D) +
L1^{n}(D) = 0, \eeq
\item Define the defect $d_\mathrm{j,k}$ at every grid point:
\beq
 d_\mathrm{j,k} = - \DD{\delta D}{\delta  t} - \vartheta
L1^{n+1}_\mathrm{high}(D) - (1-\vartheta)
L1^{n}_\mathrm{high}(D)|_{j,k},
 \eeq where the subscript "high" means  that the transport operators
are evaluated using a spatially accurate advection scheme.\\

\item Define at each grid point the following operator:
\beq LD = -\DD{\delta D}{\delta  t} - \vartheta
L1^{n+1}_\mathrm{low}(D) - (1-\vartheta)
L1^{n}_\mathrm{high}(D)|_{j,k}.
 \eeq Compute the following entries at each grid point:

\[
\underline{S}^{r}_{j,k} = \DD{\D LD}{\D D_{j-1,k}},\,\,
\mathcal{D}_{j,k} = \DD{\D LD}{\D D_{j,k}},\,\,
\overline{S}^{r}_{j,k} = \DD{\D LD}{\D D_{j+1,k}}, 
\]
\beq
\underline{S}^{z}_{j,k} = \DD{\D LD}{\D D_{j,k-1}},\,\,
\overline{S}^{z}_{j,k} = \DD{\D LD}{\D D_{j,k+1}}
\eeq

\item In the one-dimensional case, the following matrix equation should be solved at each grid point:

{\small
\beq
\left(
  \begin{array}{ccc}
    \underline{S}_{j,k}, & \mathcal{D}_{j,k}, & \overline{S}_{j,k}
  \end{array}
\right)
\left(
  \begin{array}{l}
    \delta D_{j-1,k} \\
    \delta D_{j,k} \\
    \delta D_{j+1,k} \\
  \end{array}
\right)
=
\left(
  \begin{array}{l}
    d_{j-1,k} \\
    d_{j,k} \\
    d_{j+1,k} \\
  \end{array}
\right), \textrm{ where j=1, J, and k= const.}
\eeq
} For J number of grid points in the radial direction, this
yields the tri-diagonal matrix equation:

\beq
\left(
  \begin{array}{ccccc}
    \mathsmaller{\mathcal{D}_{1}} &  \mathsmaller{{\overline{S}}_{1}} &  &  &  \\
     \mathsmaller{{\underline{S}}_{2}} & \mathsmaller{\mathcal{D}_{2}} & \mathsmaller{{\overline{S}}_{2}} &  &  \\
     & \ddots & \ddots & \ddots &  \\
      &  &   \mathsmaller{{\underline{S}}_{J}} & \mathsmaller{\mathcal{D}_{J}} & \\
  \end{array}
\right)
\left(
  \begin{array}{c}
    \mathsmaller{\delta D_{1}} \\
    \mathsmaller{\delta D_{2}}\\
    \vdots\\
    \mathsmaller{\delta D_{N}} \\
  \end{array}
\right) = \left(
  \begin{array}{c}
    \mathsmaller{d_{1}} \\
    \mathsmaller{d_{2} }\\
    \vdots\\
    \mathsmaller{d_{N}} \\
  \end{array}
\right) ~~~~\textrm{ for constant k}
 \eeq Although this matrix equation is linear in D, it should be
solved iteratively to recover the high
spatial accuracy on the right hand side. \\

Similarly, if the continuity and the radial momentum equation are to
be solved in one dimension as a coupled system, we may obtain the
following relation at each grid point:

 \beq \left(
  \begin{array}{|cc|cc|cc|}
    \mathsmaller{L11_{j-1}} & \mathsmaller{L12_{j-1}} & \mathsmaller{L11_{j}} & \mathsmaller{L12_{j}} & \mathsmaller{L11_{j+1}} & \mathsmaller{L12_{j+1}}\\
    \mathsmaller{L21_{j-1}} & \mathsmaller{L22_{j-1}} & \mathsmaller{L21_{j}} & \mathsmaller{L22_{j}} & \mathsmaller{L21_{j+1}} & \mathsmaller{L22_{j+1}}\\
  \end{array}
\right)
 \left(
  \begin{array}{l}
  \left\{
    \begin{array}{l}
    \mathsmaller{\delta D} \\
    \mathsmaller{\delta M }\\
    \end{array}
  \right\}_\mathrm{j-1} \\
   \left\{
    \begin{array}{l}
    \mathsmaller{\delta D} \\
    \mathsmaller{\delta M }\\
    \end{array}
  \right\}_\mathrm{j} \\
    \left\{
    \begin{array}{l}
    \mathsmaller{\delta D} \\
    \mathsmaller{\delta M }\\
    \end{array}
  \right\}_\mathrm{j+1}
   \end{array}
  \right)
  =
   \left(
  \begin{array}{l}
  \left\{
    \begin{array}{l}
    \mathsmaller{d^\mathrm{D}} \\
    \mathsmaller{d^\mathrm{M}}\\
    \end{array}
  \right\}_\mathrm{j-1} \\
   \left\{
    \begin{array}{l}
    \mathsmaller{d^\mathrm{D}} \\
    \mathsmaller{d^\mathrm{M}}\\
    \end{array}
  \right\}_\mathrm{j} \\
    \left\{
    \begin{array}{l}
    \mathsmaller{d^\mathrm{D}} \\
    \mathsmaller{d^\mathrm{M}}\\
    \end{array}
  \right\}_\mathrm{j+1}
   \end{array}
  \right)
  \eeq
  \[
    ~~\textrm{ for j = 1, J  and k = const.,} \]
     where $ Lmn_{l}= \DD{\D Lm}{\D q_n}|_{j=l}.$ Specifically, L1 is the density
 equation and L2 the momentum equation.\\
 For J number of points this yields a tri-diagonal block matrix,
in which each block has the dimension $2\times 2 .$\\
For a given set of equations in one dimension,   we have just to
replace the above $2\times 2$ block matrix by a square block matrix
whose dimensions are  $N_\mathrm{eq}\times N_\mathrm{eq}$, where
$N_\mathrm{eq}$ is the number of unknown variables:
\beq
\left(
  \begin{array}{cccc}
    {\mathsmaller{\square}} & {\mathsmaller{\square}} &  &  \\
    {\mathsmaller{\square}} & {\mathsmaller{\square}} & {\mathsmaller{\square}}  &  \\
     & \ddots & \ddots & \ddots \\
     &   & {\mathsmaller{\square}} & {\mathsmaller{\square}}    \\
  \end{array}
\right) \left(
  \begin{array}{l}
    \mathsmaller{\delta \overline{q}_{1}} \\
      \mathsmaller{\delta \overline{q}_{2}} \\
        \vdots \\
          \mathsmaller{\delta \overline{q}_{J}} \\
  \end{array}
\right)
=
\left(
  \begin{array}{c}
    \mathsmaller{\overline{d}_{1}} \\
    \mathsmaller{\overline{d}_{2} }\\
    \vdots\\
    \mathsmaller{\overline{d}_{J}} \\
  \end{array}
\right). \eeq
 $\overline{q}$ here is a vector of $N_\mathrm{eq}$ entries. \\

The extension into two-dimensions gives rise to a matrix equation of
the following form:
 \beq \tilde{A} \left(
  \begin{array}{l}
     \mathsmaller{\delta \overline{q}_{1,1}} \\
    \vdots \\
     \mathsmaller{\delta \overline{q}_{J,1}} \\
     \mathsmaller{\delta \overline{q}_{1,2}} \\
    \vdots \\
     \mathsmaller{\delta \overline{q}_{J,K}} \\
  \end{array}
\right) = \left(
  \begin{array}{l}
     \mathsmaller{\overline{d}_{1,1}} \\
    \vdots \\
     \mathsmaller{\overline{d}_{J,1}} \\
     \mathsmaller{\overline{d}_{1,2}} \\
    \vdots \\
     \mathsmaller{\overline{d}_{J,K}}. \\
  \end{array}
\right)
 \eeq The matrix $\tilde{A}$ has a similar structure as M1 in Fig. \ref{ClusterMatrix}.
This matrix equation is solved iteratively, using a non-direct
inversion procedure.
\een
\section{Test calculations}

\begin{figure}[htb]
\centering {
\includegraphics*[width=\hsize, bb=54 390 363 738,clip]{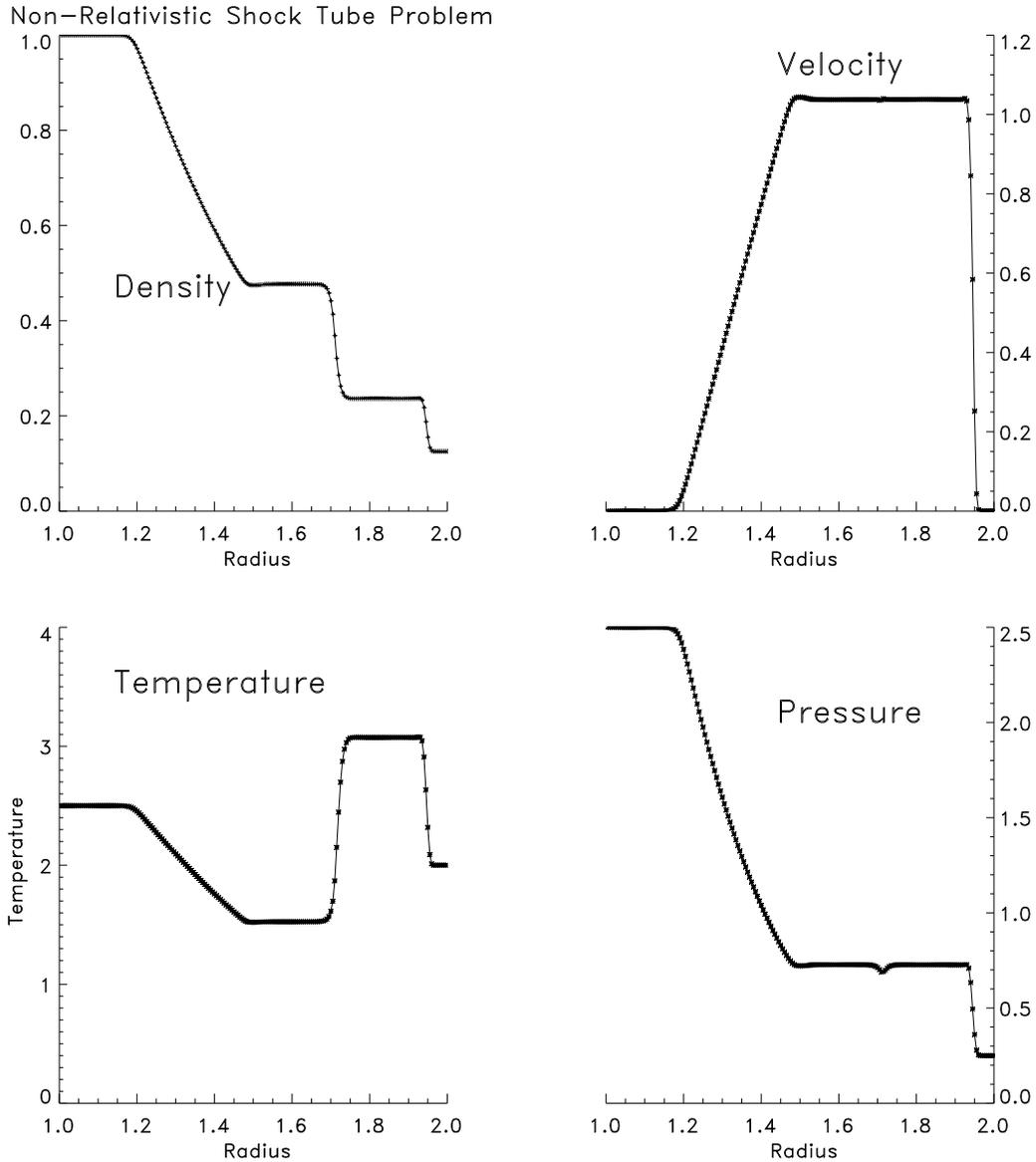}
}
\caption{\small The classical non-relativistic shock tube problem.
The profiles of the density, velocity, temperature and pressure are
displayed. The advection scheme used here is second order in time
and third order in space. 200 uniformly distributed finite volume
cells are used. } \label{NRshock}
\end{figure}

\begin{figure}[htb]
\centering {
\includegraphics*[width=\hsize, bb=45 388 370 738,clip]{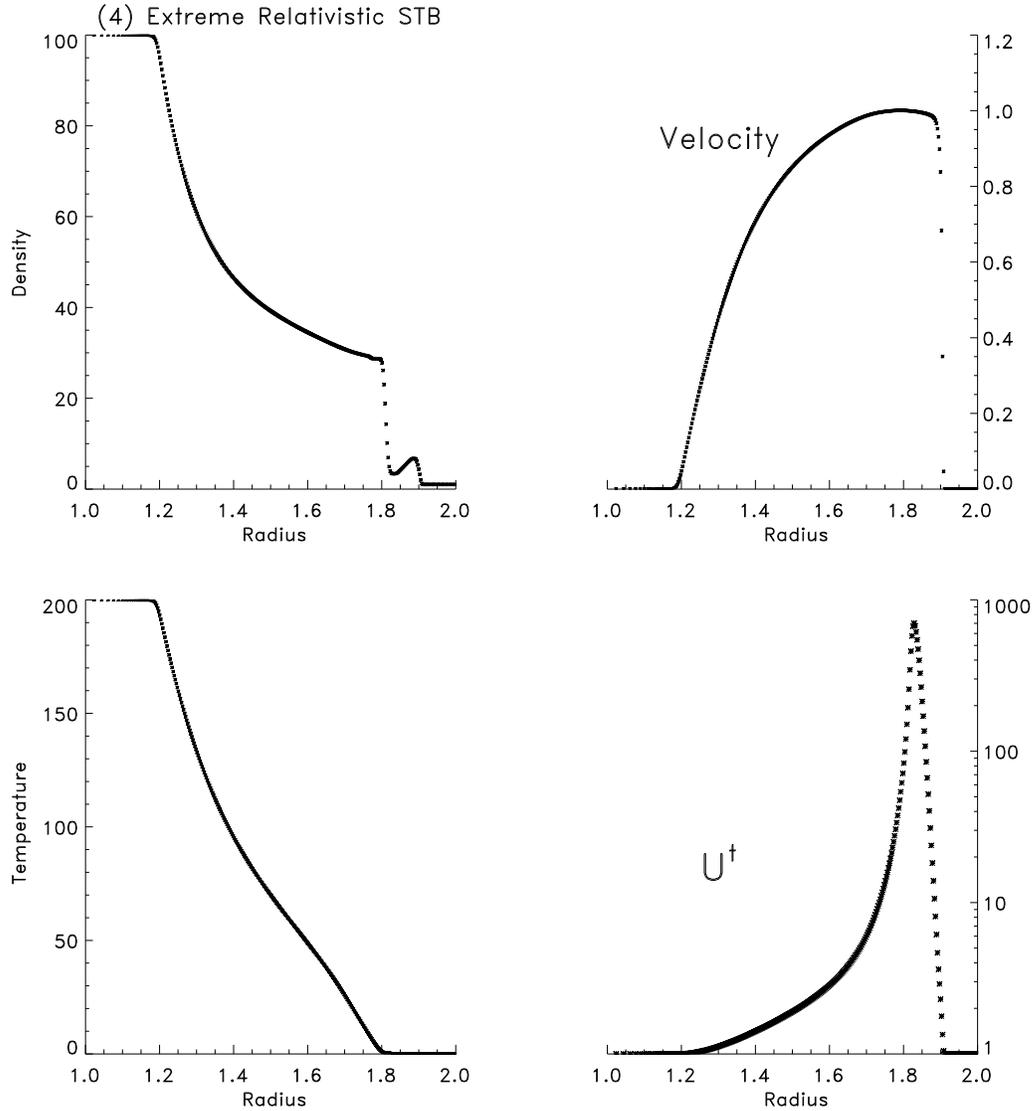}
}
\caption{\small The ultra-relativistic shock tube problem. The
radial distributions of the density, radial velocity, temperature
and the modified Lorentz factor \(U^{t}\) are shown. The accuracy of
the scheme and the number of points are identical to those in  Fig. \ref{NRshock}.
This calculation shows that shocks propagating with Lorentz
factors of order 1000 can be safely treated with our algorithm. }
\label{Rshock}
\end{figure}

The verification tests of the Newtonian version of the present
algorithm have been presented in a series of papers (see \citet{Hujeirat2005}
and the references therein). Nevertheless the modifications
made here are serious and deserve appropriate test calculations to
ensure bug-free runs as
well as a consistent re-production of the results in the Newtonian regime.\\
In the following we briefly mention several of the test calculations
performed:

\bit
  \item {\bf The shock tube problem - STP} \\
        In the case of low fluid-velocities, the modification made
        should enable capturing of shocks propagating at
        sub-relativistic speeds, irrespective of the accuracies
        used. Therefore, we have applied the algorithm to the
        well-known Sod-problem (see \citet{Hujeirat1995} and the references
        therein). Fig. \ref{NRshock} shows that the algorithm is indeed
        capable of re-producing Sod's solution with high accuracy.
  \item {\bf The ultra-relativistic shock tube problem} \\
        The speed of the shocks in the Sod's problem  can be  made
        arbitrary large, depending on the initial ratio of the
        pressure in the tube. While  non-relativistic solvers may
        produce propagating velocities that exceeds the speed of
        light, a conservative and accurate relativistic solver
        should produce velocities that can be extremely close to but
        never exceed the speed of light.\\
        {In Fig. \ref{Rshock} the one-dimensional profiles of
        the density, velocity, temperature and Lorentz factor $u^\mm{t}$
        are displayed. These profiles agree qualitatively with the analytical solution
        of the relativistic STP provided by \cite{MartiMueller2003}. In a forthcoming paper, we intend
        to quantitatively compare the profiles for extremely large Lorentz factors. }

        Fig. \ref{Rshock} demonstrates the
        strong robustness of the algorithm and its capability to
        capture the propagation of extreme ultra-relativistic shocks
        in which the Lorentz factor is of order 1000. Such
        robustness is essential to enable modeling jetted Gamma-Ray
        bursts, where the Lorentz factors are in the excess of
        several hundreds.

\begin{figure}[htb]
 \centering {
 \includegraphics*[width=0.8 \hsize, bb=54 215 380 738,clip]{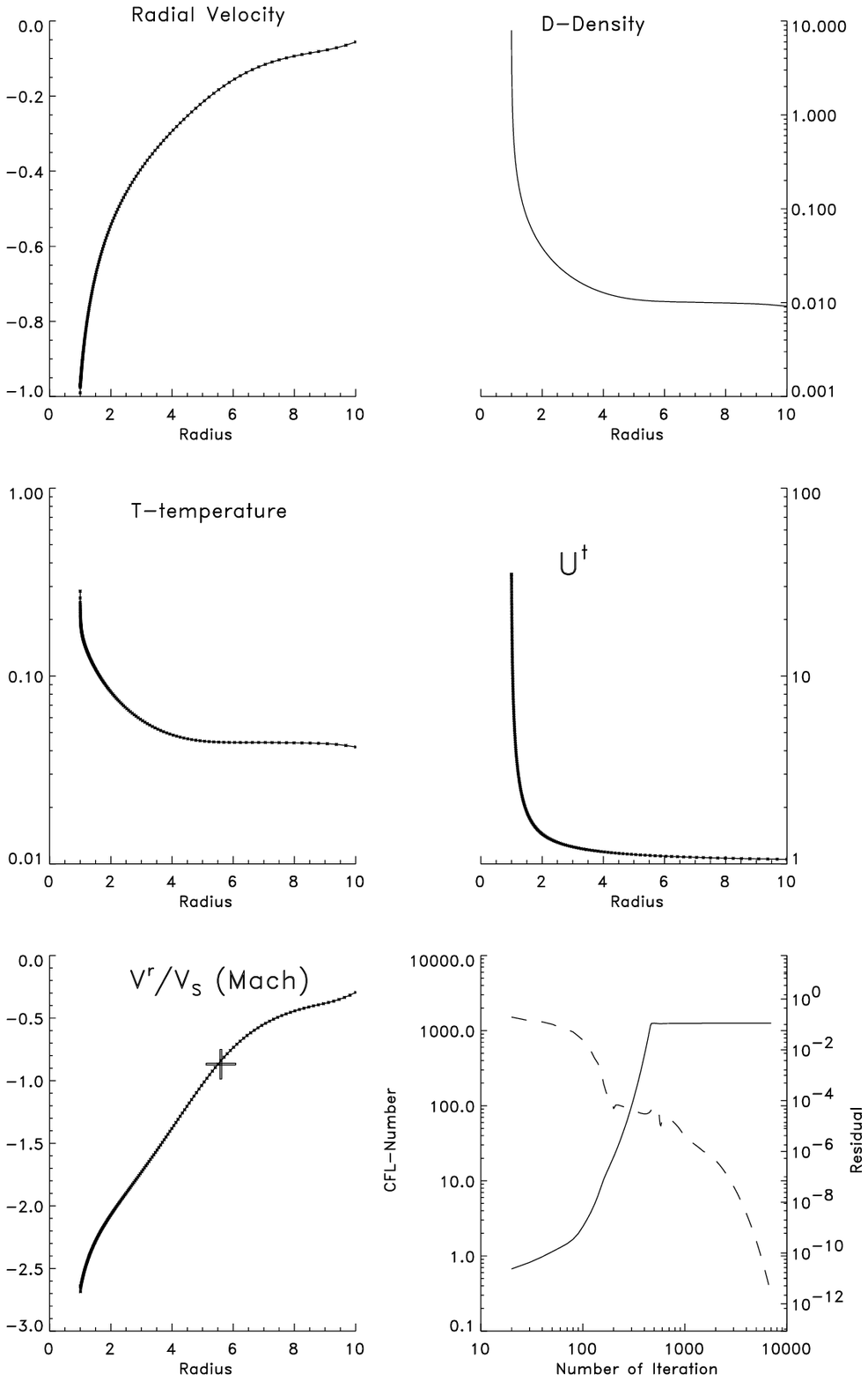}
 }
 \caption{\small The Bondi accretion problem onto a Schwarzschild
 black hole. The profiles of the radial velocity, the relativistic
 density, temperature, the Lorentz factor \(U^t\) and the radial
 Mach number. In the lower-right panel profiles of the Courant
 number (solid line) and the corresponding residual (dashed line)
 versus the iteration number
 are displayed. Although the problem is spherically symmetric, the
 calculations have been carried out using 200 grid points in the
 radial and 30 in the horizontal-direction. The accuracy of the
 advection scheme is set to be first order in time and third order
 in space. This test enables us to examine the capability of the
 algorithm at  capturing steady solutions that are essentially
 one-dimensional using a 2D numerical scheme. We have verified that
 the 30-profiles in the radial direction obtained at different
 $\theta$ are identical to machine- accuracy.  } \label{Bondi}
\end{figure}

  \item {\bf Relativistic Bondi accretion onto Schwarzschild black holes }  \\
  This problem is appropriate for testing the capability of the
  solver at treating transonic stationary accretion flows onto
  Schwarzschild black holes, assuming perfect spherical symmetry.
  This problem has been investigated by several authors (see \citet{Michel1972},
  also see \citet{Hawley_etal1984a, Hawley_etal1984b} for a comprehensive description of the
  numerical treatment). In this problem, a constant flux of an ideal
  gas is said to be accreted by a non-rotating black hole. Depending
  on location of the outer boundary and on the temperature of the
  flow,
   the initially subsonic inflow should make a transition
   into the super-sonic regime at a specific
    radius, which appears to be determined entirely by the constant
   of motion. On the other hand, the Lorentz factor of the flow as
   it crosses the inner boundary should approach the speed of light,
   depending on how close the inner radius is to the event horizon. In
   Fig. \ref{Bondi} we display the radial distributions of the velocity,
   density, temperature, Lorentz factor and the Mach number, which
   clearly well-agree with the known analytical solutions. In
   obtaining these results we used a pseudo time-stepping scheme to
   enhance convergence. The very last time step size in this
   calculations corresponds to Courant number 2000, approximately.

  \item {\bf Standing shocks around black holes}\\
The purpose of this test is mainly to examine the capability of the
algorithm  at re-producing the formation of the two-dimensional
curved standing shocks around a Schwarzschild black hole that have
been obtained using the Newtonian version of the algorithm.

\begin{figure}[hbt]
\begin{center}
{
\includegraphics*[width=0.43 \hsize,bb=40 262 245 487,clip]{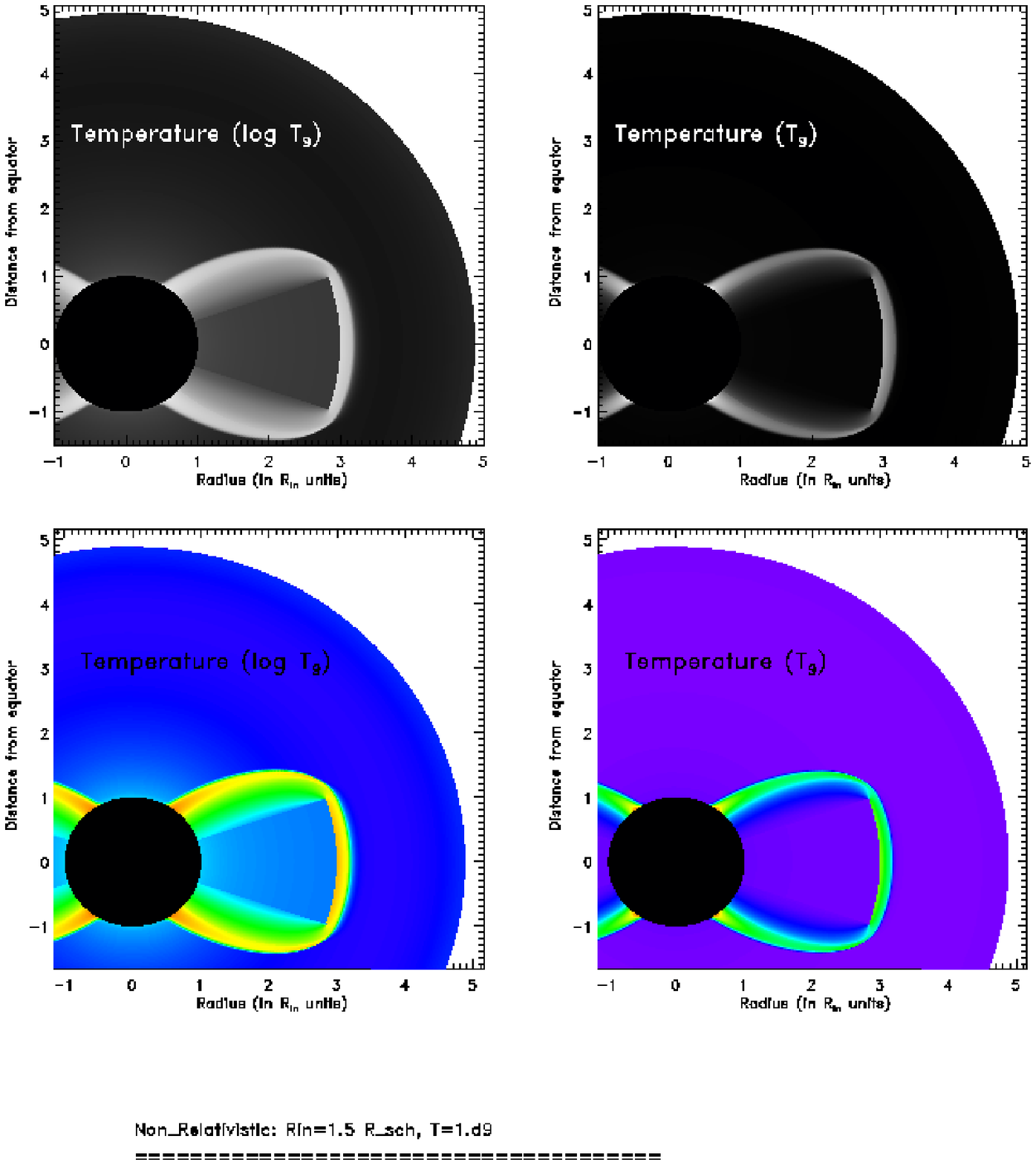} \\
\includegraphics*[width=0.43 \hsize,bb=40 262 245 487,clip]{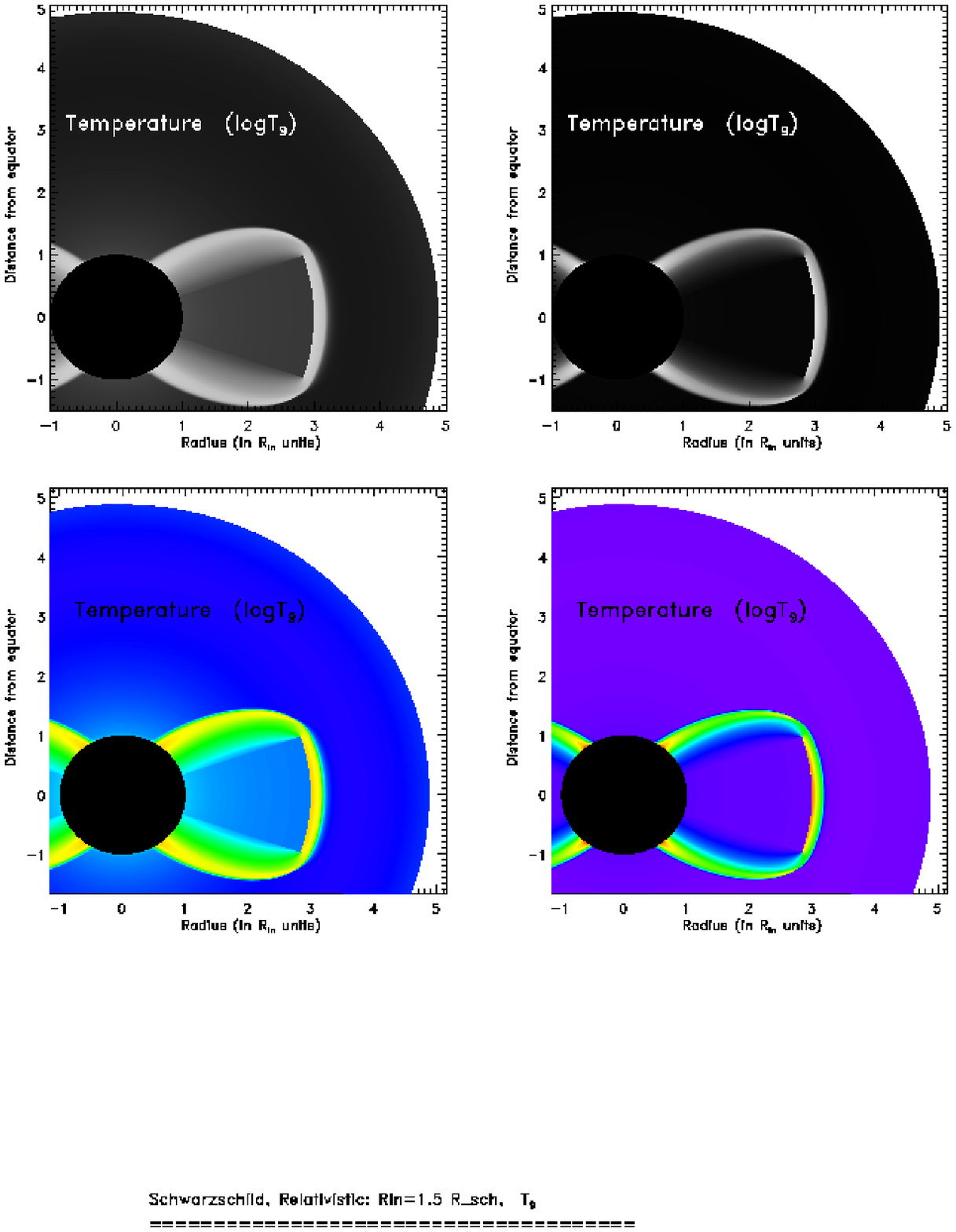} \\
\includegraphics*[width=0.43 \hsize,bb=40 252 245 487,clip]{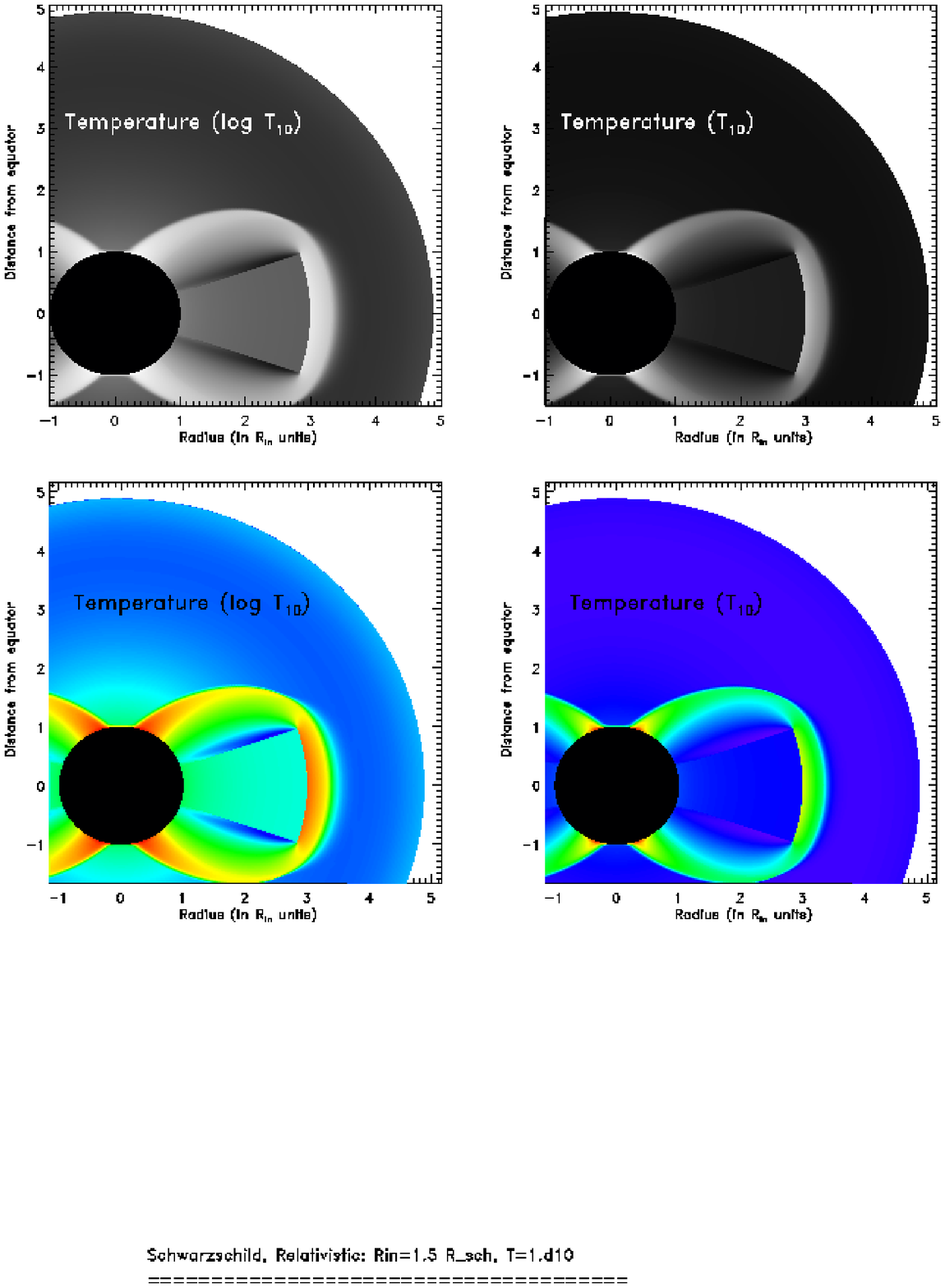} \\

}
\end{center}
{\vspace*{-0.0cm}} \caption{\small The 2D distribution of the
temperature (in units of $10^9$K) of a freely falling
non-relativistic gas onto a Schwarzschild black hole surrounded by a
static cold disk (top panel). In this figure, color gradients run as
follows: red color corresponds to large temperature-values, green to
intermediate and blue to low values. The distribution in the second
and third panels have been obtained using the general relativistic
version of the algorithm. Here the inflowing matter across the outer
boundary has the temperatures  $10^9$K (middle) and $10^{10}$K
(bottom). } \label{FFS2D}
\end{figure}

\begin{figure}[hbt]
\begin{center}
{
\includegraphics*[width=0.3 \hsize,bb=67 508 270 724,clip]{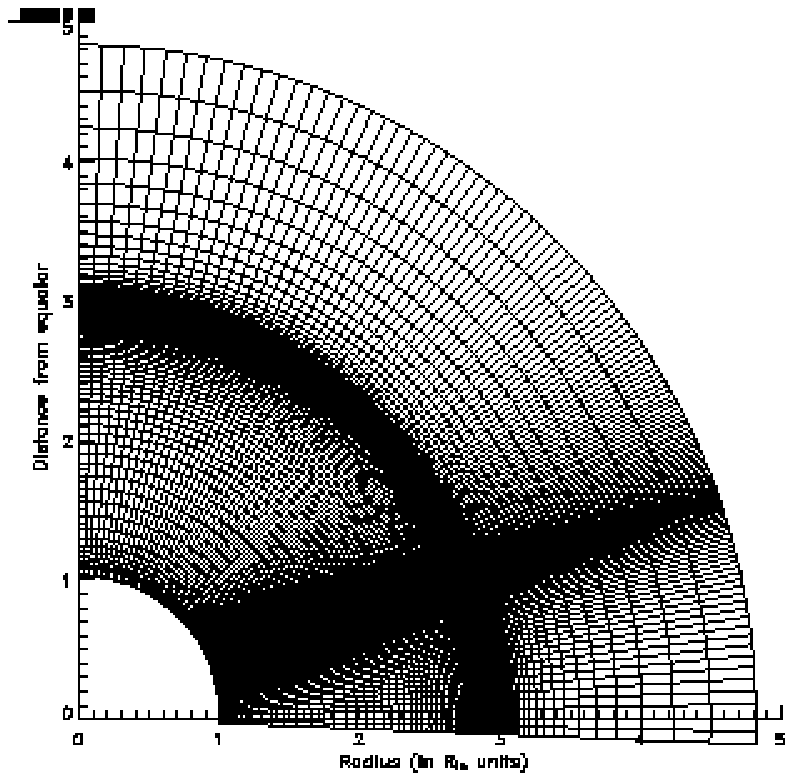}\\
\includegraphics*[width=0.6 \hsize,bb=136 130 403 636,clip]{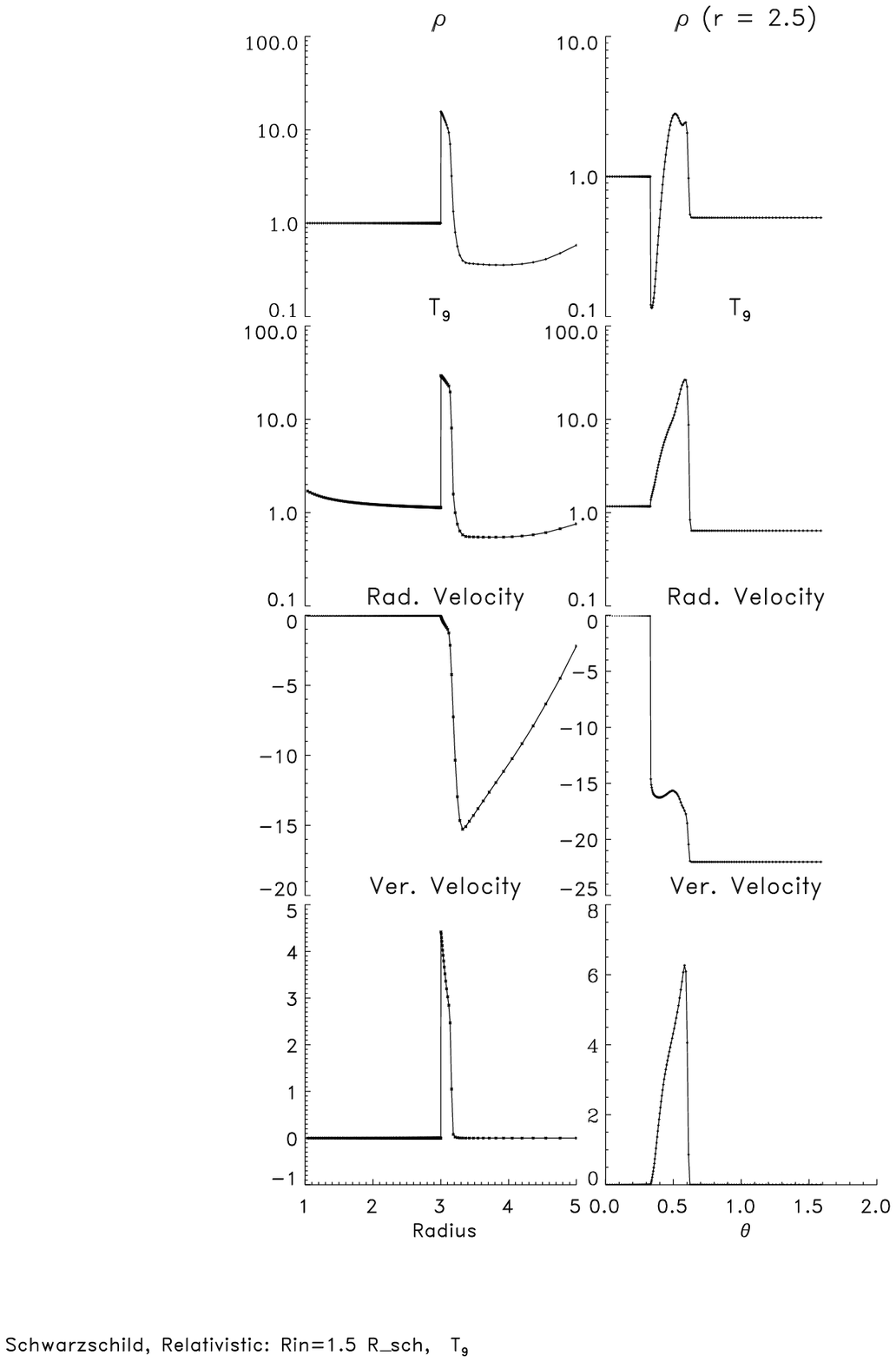}
}
\end{center}
{\vspace*{-0.0cm}} \caption{\small Distribution of the grid points
used in the calculations. A strong non-uniform distribution of the
grid points has been constructed to enable accurate capturing  of
standing shocks surrounding the cold disk. The tensor-product mesh
consists of 275 finite volume cells in the radial and 130 in the
horizontal direction, respectively. In the lower panel the profiles
of the density, temperature, radial and vertical velocities along
the equator and horizontal along the constant radius $r=2.5$ are
shown. } \label{FFS1D}
\end{figure}

This problem is similar to the forward facing step in computational
fluid dynamics. Here a cold and dense disk has been placed in the
innermost equatorial region: $[1\le r\le 10]\times [-0.3 \le \theta
\le 0.3]$ (see Figures \ref{FFS2D} and \ref{FFS1D}). Vanishing in- and out-flow
conditions have been imposed at the boundaries of the cold disk. The
gas surrounding the disk is taken to be inviscid, thin, hot and
non-rotating. The cold disk here serves as a two-dimensional barrier
that disturbs the gas from otherwise a spherically symmetric freely
falling flow onto a Schwarzschild black hole and, instead, it forces
the inflow to form a curved shock which eclipses the cold disk.

In solving the HD-equations, an advection scheme  of third order
spatial accuracy and of first order accuracy in time has been used.

Hence the scheme is taken to be highly diffusive in time in order to
damp oscillations and to accelerate convergence into steady-state.
The domain of calculation is sub-divided into 200 strongly-stretched
finite volume cells in the radial and 60 in the horizontal
direction. In Fig. \ref{FFS1D} the 1D radial and horizontal profiles, the 2D
configuration of the density, temperature and the velocity field are
shown. Indeed, the algorithm shows that it is numerically stable and
capable of capturing steady-state shocks with complicated shock
structures even for large CFL-numbers. 

\eit


\section{Summary}
In this paper we have extended the previous Newtonian implicit
algorithm to enable solving the hydrodynamical equations in general
relativity. The 3D axi-symmetric hydrodynamical equations have been
presented in the background of a Kerr metric of a black hole using
the Boyer-Lindquist coordinates. The equations have been formulated
in conservative form and subsequently solved numerically, using the
finite volume formulation. The new extension can be well
accommodated within the hierarchical solution scenario, in which the
degree of implicitness can be made dynamical, depending on the
hydrodynamical problem in hand. In particular, for modeling strongly
time-dependent astrophysical flows, such as moving shocks, the
pre-conditioners used are tri-diagonal matrices that are solved
successively. Although the computational costs per time step may be
one order of magnitude larger than their explicit counterparts, this
can be compensated through a reduction of the overall number of time
steps
required to recover a physically reliable time scale. 

On the other hand, the efficiency and robustness of the HSS are
superior, if the solutions sought are stationary or
quasi-stationary, irrespective of whether the flow is dissipative or
not.

Finally, a unification scheme for various numerical methods has been
presented. In particular, the HSS algorithm enables the construction
of a large variety of solvers, in which the degree of implicitness
may range from purely explicit up to strongly implicit, depending on
the physical properties of the underlying flow problem. Thus, the
HSS is actually a unified algorithm for treating weakly
compressible, incompressible, time-dependent, time-independent,
radiative, magnetized non-dissi\-pative or strongly dissipative flows.
As a consequence, using the HSS algorithm, we are able to save a
large number of working hours which otherwise would go in designing
different solvers for different physical problems.\\
In a subsequent paper, we intend to discuss and describe the
inclusion of the magnetohydrodynamical equations in general
relativity into the present solver. 

{\bf Acknowledgment} A.H. thanks Prof. J. Du\v{s}ek for
reading the manuscript and for the hospitality during his visit to
the Institut de mechanique des fluides et des solides, CNRS and the
Louis Pasteur University in Strasbourg.
This work has been partially supported by the Klaus-Tschira Stiftung under the
project number 00.099.2006.



\end{document}